\documentclass[longauth]{aa}  

\usepackage{graphicx}
\usepackage{txfonts}
\usepackage{hyperref}
\usepackage{euclid}
\usepackage{bm}
\usepackage[dvipsnames]{xcolor}
  
\usepackage[T1]{fontenc} 
\usepackage{braket}

\newcommand{\GCs}{\text{GC}\ensuremath{_\mathrm{s}}}
\newcommand{\GCph}{\text{GC}\ensuremath{_\mathrm{ph}}}

\newcommand{\Omegam}{\Omega_{\mathrm{m},0}}
\newcommand{\Omegab}{\Omega_{\mathrm{b},0}}

\renewcommand\aap{A\&A}

\defcitealias{IST:paper1}{EC19}

\usepackage[nameinlink,noabbrev]{cleveref}
\Crefname{equation}{Eq.}{Eqs.}
\Crefname{section}{Sect.}{Sects.}
\Crefname{figure}{Fig.}{Figs.}
\crefname{equation}{Equation}{Equations}
\crefname{section}{Section}{Sections}
\crefname{figure}{Figure}{Figures}

\begin{document} 


\title{\Euclid: The importance of galaxy clustering and weak lensing cross-correlations within the photometric \Euclid survey\thanks{This paper is published on behalf of the Euclid Consortium.}}


\author{I.~Tutusaus$^{1,2,3}$\thanks{\email{tutusaus@ice.csic.es}}, M.~Martinelli$^{4}$, V.F.~Cardone$^{5,6}$, S.~Camera$^{7,8,9}$, S.~Yahia-Cherif$^{2}$, S.~Casas$^{10}$, A.~Blanchard$^{2}$, M.~Kilbinger$^{10,11}$, F.~Lacasa$^{12,13}$, Z.~Sakr$^{2,14}$, S.~Ili\'c$^{2,15,16}$, M.~Kunz$^{17}$, C.~Carbone$^{18}$, F.J.~Castander$^{1,3}$, F.~Dournac$^{2}$, P.~Fosalba$^{1,3}$, T.~Kitching$^{19}$, K.~Markovic$^{20}$, A.~Mangilli$^{2}$, V.~Pettorino$^{21}$, D.~Sapone$^{22}$, V.~Yankelevich$^{23}$, N.~Auricchio$^{24}$, R.~Bender$^{25,26}$, D.~Bonino$^{9}$, A.~Boucaud$^{27}$, M.~Brescia$^{28}$, V.~Capobianco$^{9}$, J.~Carretero$^{29}$, M.~Castellano$^{6}$, S.~Cavuoti$^{28,30,31}$, R.~Cledassou$^{12}$, G.~Congedo$^{32}$, L.~Conversi$^{33,34}$, L.~Corcione$^{9}$, A.~Costille$^{35}$, M.~Cropper$^{19}$, F.~Dubath$^{36}$, S.~Dusini$^{37}$, G.~Fabbian$^{38}$, M.~Frailis$^{39}$, E.~Franceschi$^{24}$, B.~Garilli$^{18}$, F.~Grupp$^{26}$, L.~Guzzo$^{40,41,42}$, H.~Hoekstra$^{43}$, F.~Hormuth$^{44}$, H.~Israel$^{25}$, K.~Jahnke$^{45}$, S.~Kermiche$^{46}$, B.~Kubik$^{47}$, R.~Laureijs$^{48}$, S.~Ligori$^{9}$, P.B.~Lilje$^{49}$, I.~Lloro$^{50}$, E.~Maiorano$^{24}$, O.~Marggraf$^{51}$, R.~Massey$^{52}$, S.~Mei$^{53}$, E.~Merlin$^{6}$, G.~Meylan$^{54}$, L.~Moscardini$^{24,55,56}$, P.~Ntelis$^{46}$, C.~Padilla$^{29}$, S.~Paltani$^{36}$, F.~Pasian$^{39}$, W.J.~Percival$^{57,58,59}$, S.~Pires$^{10}$, M.~Poncet$^{12}$, F.~Raison$^{26}$, J.~Rhodes$^{20}$, M.~Roncarelli$^{24,55}$, E.~Rossetti$^{55}$, R.~Saglia$^{25,26}$, P.~Schneider$^{51}$, A.~Secroun$^{46}$, S.~Serrano$^{1,3}$, C.~Sirignano$^{37,60}$, G.~Sirri$^{61}$, J.~Starck$^{10}$, F.~Sureau$^{10}$, A.N.~Taylor$^{32}$, I.~Tereno$^{62,63}$, R.~Toledo-Moreo$^{64}$, L.~Valenziano$^{24,61}$, Y.~Wang$^{65}$, N.~Welikala$^{32}$, J.~Weller$^{25,26}$, A.~Zacchei$^{39}$, J.~Zoubian$^{46}$}

\institute{$^{1}$ Institute of Space Sciences (ICE, CSIC), Campus UAB, Carrer de Can Magrans, s/n, 08193 Barcelona, Spain\\
$^{2}$ Institut de Recherche en Astrophysique et Plan\'etologie (IRAP), Universit\'e de Toulouse, CNRS, UPS, CNES, 14 Av. Edouard Belin, F-31400 Toulouse, France\\
$^{3}$ Institut d’Estudis Espacials de Catalunya (IEEC), 08034 Barcelona, Spain\\
$^{4}$ Instituto de F\'isica T\'eorica UAM-CSIC, Campus de Cantoblanco, E-28049 Madrid, Spain\\
$^{5}$ I.N.F.N.-Sezione di Roma Piazzale Aldo Moro, 2 - c/o Dipartimento di Fisica, Edificio G. Marconi, I-00185 Roma, Italy\\
$^{6}$ INAF-Osservatorio Astronomico di Roma, Via Frascati 33, I-00078 Monteporzio Catone, Italy\\
$^{7}$ INFN-Sezione di Torino, Via P. Giuria 1, I-10125 Torino, Italy\\
$^{8}$ Dipartimento di Fisica, Universit\'a degli Studi di Torino, Via P. Giuria 1, I-10125 Torino, Italy\\
$^{9}$ INAF-Osservatorio Astrofisico di Torino, Via Osservatorio 20, I-10025 Pino Torinese (TO), Italy\\
$^{10}$ AIM, CEA, CNRS, Universit\'{e} Paris-Saclay, Universit\'{e} Paris Diderot, Sorbonne Paris Cit\'{e}, F-91191 Gif-sur-Yvette, France\\
$^{11}$ Institut d'Astrophysique de Paris, 98bis Boulevard Arago, F-75014, Paris, France\\
$^{12}$ Centre National d'Etudes Spatiales, Toulouse, France\\
$^{13}$ Institut d'Astrophysique Spatiale, Universite Paris-Sud, Batiment 121, 91405 Orsay, France\\
$^{14}$ Universit\'e St Joseph; UR EGFEM, Faculty of Sciences, Beirut, Lebanon\\
$^{15}$ Universit\'e PSL, Observatoire de Paris, Sorbonne Universit\'e, CNRS, LERMA, F-75014, Paris, France\\
$^{16}$ CEICO, Institute of Physics of the Czech Academy of Sciences, Na Slovance 2, Praha 8, Czech Republic\\
$^{17}$ Universit\'e de Gen\`eve, D\'epartement de Physique Th\'eorique and Centre for Astroparticle Physics, 24 quai Ernest-Ansermet, CH-1211 Gen\`eve 4, Switzerland\\
$^{18}$ INAF-IASF Milano, Via Alfonso Corti 12, I-20133 Milano, Italy\\
$^{19}$ Mullard Space Science Laboratory, University College London, Holmbury St Mary, Dorking, Surrey RH5 6NT, UK\\
$^{20}$ Jet Propulsion Laboratory, California Institute of Technology, 4800 Oak Grove Drive, Pasadena, CA, 91109, USA\\
$^{21}$ CEA Saclay, DFR/IRFU, Service d'Astrophysique, Bat. 709, 91191 Gif-sur-Yvette, France\\
$^{22}$ Departamento de F\'isica, FCFM, Universidad de Chile, Blanco Encalada 2008, Santiago, Chile\\
$^{23}$ Astrophysics Research Institute, Liverpool John Moores University, 146 Brownlow Hill, Liverpool L3 5RF, UK\\
$^{24}$ INAF-Osservatorio di Astrofisica e Scienza dello Spazio di Bologna, Via Piero Gobetti 93/3, I-40129 Bologna, Italy\\
$^{25}$ Universit\"ats-Sternwarte M\"unchen, Fakult\"at f\"ur Physik, Ludwig-Maximilians-Universit\"at M\"unchen, Scheinerstrasse 1, 81679 M\"unchen, Germany\\
$^{26}$ Max Planck Institute for Extraterrestrial Physics, Giessenbachstr. 1, D-85748 Garching, Germany\\
$^{27}$ APC, AstroParticule et Cosmologie, Universit\'e Paris Diderot, CNRS/IN2P3, CEA/lrfu, Observatoire de Paris, Sorbonne Paris Cit\'e, 10 rue Alice Domon et L\'eonie Duquet, 75205, Paris Cedex 13, France\\
$^{28}$ INAF-Osservatorio Astronomico di Capodimonte, Via Moiariello 16, I-80131 Napoli, Italy\\
$^{29}$ Institut de F\'isica d’Altes Energies IFAE, 08193 Bellaterra, Barcelona, Spain\\
$^{30}$ Department of Physics "E. Pancini", University Federico II, Via Cinthia 6, I-80126, Napoli, Italy\\
$^{31}$ INFN section of Naples, Via Cinthia 6, I-80126, Napoli, Italy\\
$^{32}$ Institute for Astronomy, University of Edinburgh, Royal Observatory, Blackford Hill, Edinburgh EH9 3HJ, UK\\
$^{33}$ European Space Agency/ESRIN, Largo Galileo Galilei 1, 00044 Frascati, Roma, Italy\\
$^{34}$ ESAC/ESA, Camino Bajo del Castillo, s/n., Urb. Villafranca del Castillo, 28692 Villanueva de la Ca\~nada, Madrid, Spain\\
$^{35}$ Aix-Marseille Univ, CNRS, CNES, LAM, Marseille, France\\
$^{36}$ Department of Astronomy, University of Geneva, ch. d'\'Ecogia 16, CH-1290 Versoix, Switzerland\\
$^{37}$ INFN-Padova, Via Marzolo 8, I-35131 Padova, Italy\\
$^{38}$ Department of Physics \& Astronomy, University of Sussex, Brighton BN1 9QH, UK\\
$^{39}$ INAF-Osservatorio Astronomico di Trieste, Via G. B. Tiepolo 11, I-34131 Trieste, Italy\\
$^{40}$ Dipartimento di Fisica "Aldo Pontremoli", Universit\'a degli Studi di Milano, Via Celoria 16, I-20133 Milano, Italy\\
$^{41}$ INFN-Sezione di Milano, Via Celoria 16, I-20133 Milano, Italy\\
$^{42}$ INAF-Osservatorio Astronomico di Brera, Via Brera 28, I-20122 Milano, Italy\\
$^{43}$ Leiden Observatory, Leiden University, Niels Bohrweg 2, 2333 CA Leiden, The Netherlands\\
$^{44}$ von Hoerner \& Sulger GmbH, Schlo{\ss}Platz 8, D-68723 Schwetzingen, Germany\\
$^{45}$ Max-Planck-Institut f\"ur Astronomie, K\"onigstuhl 17, D-69117 Heidelberg, Germany\\
$^{46}$ Aix-Marseille Univ, CNRS/IN2P3, CPPM, Marseille, France\\
$^{47}$ Institut de Physique Nucl\'eaire de Lyon, 4, rue Enrico Fermi, 69622, Villeurbanne cedex, France\\
$^{48}$ European Space Agency/ESTEC, Keplerlaan 1, 2201 AZ Noordwijk, The Netherlands\\
$^{49}$ Institute of Theoretical Astrophysics, University of Oslo, P.O. Box 1029 Blindern, N-0315 Oslo, Norway\\
$^{50}$ NOVA optical infrared instrumentation group at ASTRON, Oude Hoogeveensedijk 4, 7991PD, Dwingeloo, The Netherlands\\
$^{51}$ Argelander-Institut f\"ur Astronomie, Universit\"at Bonn, Auf dem H\"ugel 71, 53121 Bonn, Germany\\
$^{52}$ Institute for Computational Cosmology, Department of Physics, Durham University, South Road, Durham, DH1 3LE, UK\\
$^{53}$ Universit\'{e} de Paris, F-75013, Paris, France, LERMA, Observatoire de Paris, PSL Research University, CNRS, Sorbonne Universit\'e, F-75014 Paris, France\\
$^{54}$ Observatoire de Sauverny, Ecole Polytechnique F\'ed\'erale de Lau- sanne, CH-1290 Versoix, Switzerland\\
$^{55}$ Dipartimento di Fisica e Astronomia, Universit\'a di Bologna, Via Gobetti 93/2, I-40129 Bologna, Italy\\
$^{56}$ INFN-Bologna, Via Irnerio 46, I-40126 Bologna, Italy\\
$^{57}$ Perimeter Institute for Theoretical Physics, Waterloo, Ontario N2L 2Y5, Canada\\
$^{58}$ Department of Physics and Astronomy, University of Waterloo, Waterloo, Ontario N2L 3G1, Canada\\
$^{59}$ Centre for Astrophysics, University of Waterloo, Waterloo, Ontario N2L 3G1, Canada\\
$^{60}$ Dipartimento di Fisica e Astronomia “G.Galilei", Universit\'a di Padova, Via Marzolo 8, I-35131 Padova, Italy\\
$^{61}$ INFN-Sezione di Bologna, Viale Berti Pichat 6/2, I-40127 Bologna, Italy\\
$^{62}$ Instituto de Astrof\'isica e Ci\^encias do Espa\c{c}o, Faculdade de Ci\^encias, Universidade de Lisboa, Tapada da Ajuda, PT-1349-018 Lisboa, Portugal\\
$^{63}$ Departamento de F\'isica, Faculdade de Ci\^encias, Universidade de Lisboa, Edif\'icio C8, Campo Grande, PT1749-016 Lisboa, Portugal\\
$^{64}$ Universidad Polit\'ecnica de Cartagena, Departamento de Electr\'onica y Tecnolog\'ia de Computadoras, 30202 Cartagena, Spain\\
$^{65}$ Infrared Processing and Analysis Center, California Institute of Technology, Pasadena, CA 91125, USA\\
}

\date{}

\authorrunning{I. Tutusaus et al.}

\titlerunning{The importance of galaxy clustering and weak lensing cross-correlations within the photometric \Euclid survey}

 
  \abstract
   {The data from the \Euclid mission will enable the measurement of the photometric redshifts, angular positions, and weak lensing shapes for over a billion galaxies. This large dataset, with well controlled systematic effects, will allow for cosmological analyses using the angular clustering of galaxies (\GCph) and cosmic shear (WL). For \Euclid these two cosmological probes will not be independent because they will probe the same volume of the Universe. The cross-correlation (XC) between these probes can tighten constraints and it is therefore important to quantify their impact for \Euclid.}
   {In this study we therefore extend the recent \Euclid forecasts presented in \citet{IST:paper1} by carefully quantifying the impact of XC not only on the final parameter constraints for different cosmological models, but also on the nuisance parameters. In particular, we aim at understanding the amount of additional information that XC can provide for parameters encoding systematic effects, such as galaxy bias or intrinsic alignments (IA).}
   {We follow the Fisher matrix formalism presented in \citet{IST:paper1} and make use of the codes validated therein. We also investigate a different galaxy bias model, obtained from the Flagship simulation, and additional photometric-redshift uncertainties, and the impact of including the XC terms in constraining these.}
   {Starting with a baseline model, we show that the XC terms improve the dark energy Figure of Merit (FoM) by a factor $\sim 5$, whilst they also reduce the uncertainties on galaxy bias by $\sim 17\%$ and the uncertainties on IA by a factor $\sim 4$. 
   The XC terms also help in constraining the $\gamma$ parameter for minimal modified gravity models. Concerning galaxy bias, we observe that the role of the XC terms on the final parameter constraints is qualitatively the same irrespective of the specific galaxy bias model used. For IA we show that the XC terms can help in distinguishing between different models, and that if IA terms are neglected then this can lead to significant biases on the cosmological parameters. Finally, we show that the XC terms can lead to a better determination of the mean of the photometric galaxy distributions.}
   {We find that the XC between \GCph\ and WL within the \Euclid survey is necessary to extract the full information content from the data in future analyses. These terms help in better constraining the cosmological model, and also lead to a better understanding of the systematic effects that contaminate these probes. Furthermore, we find that XC significantly helps in constraining the mean of the photometric-redshift distributions, but, at the same time, it requires a more precise knowledge of this mean, with respect to single probes, in order not to degrade the final FoM.}


   \maketitle


\section{Introduction}
To better understand the source of cosmic acceleration and the physics of gravity on cosmological scales, large galaxy surveys rely on two main probes: galaxy positions (including redshifts) and weak lensing shapes. Galaxy clustering probes the fluctuations of the underlying dark matter density and velocity fields from the angular and radial positions of galaxies. This can be used for cosmological constraints as it encodes geometric information such as the baryon acoustic oscillations \citep[BAO;][]{2005ApJ...633..560E,2015PhRvD..92l3516A}, growth information from redshift-space distortions \citep[RSD;][]{2009MNRAS.393..297P}, as well as more detailed and model-dependent cosmological information encoded in the full shape of the power spectrum  \citep{2006MNRAS.366..189S,2010MNRAS.404...60R}. Similarly, the statistical properties of large ensembles of galaxy shapes can be used to reveal the tiny signal of distortions in their images due to the gravitational potential wells traversed by photons in their propagation towards us -- a weak gravitational lensing signal known as `cosmic shear' \citep[see e.g.][for a recent review]{Kilbinger_review}. This is sensitive to the total amount of matter in the Universe and to the amplitude of its fluctuations, as well as the physics of the gravitational interaction.

Galaxy clustering and cosmic shear are the main probes of the cosmological community's scientific program for future Stage-IV experiments \citep{detf}. In this article we are interested in forecasting the capability of one such future survey: the upcoming European Space Agency (ESA) satellite \Euclid \citep{2011arXiv1110.3193L}, whose characteristics will be summarised in \Cref{sec:Euclid-survey}. To extract the full information content from \Euclid several systematic effects will have to be overcome. Some of the more important systematic effects that may affect the cosmological analysis are: the modeling of galaxy bias and redshift-space distortions, photometric-redshift uncertainties and biases, and galaxy intrinsic alignments \citep[IAs;][]{iareview1}.

Here we focus on quantifying the additional information that can be obtained through the cross-correlation (XC) of the angular power spectra of the weak lensing cosmic shear (WL) and galaxy clustering from the photometric sample (\GCph) in \Euclid. In a previous work \citep[][from hereafter EC19]{IST:paper1}, it has been shown that the combination of the information from WL, \GCph\ and their XC can result in a significant enhancement to the {forecast} Figure of Merit (FoM) on dark energy. The importance of combining WL and \GCph\ to test cosmological models and modified gravity has also been previously highlighted by several authors \citep[e.g.][]{2007PhRvL..99n1302Z,2009JCAP...10..004S,2010PhRvD..81b3503G,2010Natur.464..256R,2012MNRAS.422.2904G,2015MNRAS.451.1553E,2015MNRAS.452.2149E,2015MNRAS.452.2168E,Fonseca:2015laa,2016MNRAS.456.2806B,10.1093/mnras/sty2168}. Furthermore, as shown by \citet{Camera:2016owj,Harrison:2016stv}, the XC terms can greatly improve understanding of systematic effects. 

Other collaborations which have now released their first cosmological results from recent data, such as DES \citep{DESY1-WL, DESY1-GCWL} and KiDS+GAMA \citep{ Hildebrandt:2017qln, vanUitert:2017ieu} have also used the power of the joint analysis to improve their constraints.
For instance, the DES Collaboration uses the auto- and cross-correlations of two galaxy catalogs. The first catalog contains the positions of the lens galaxies used for galaxy-galaxy lensing and \GCph\ measurements, while the second catalog contains the positions and shape measurements of the galaxies used in the WL analysis, which also serve as source galaxies for galaxy-galaxy lensing. The DES Collaboration \citep[see details of the modeling in][]{Krause:2017ekm} works with a data vector that contains the different two-point correlation functions in real space, which in the flat-sky approximation are spherical Bessel integrals over the angular power spectra used in \citetalias{IST:paper1}. Since there are three two-point correlation functions (one for \GCph, one for WL, and one for galaxy-galaxy lensing), this joint analysis is also known in the cosmological community as a 3x2pt analysis. This kind of analysis can even be extended by including information from the cosmic microwave background, which leads to a 6x2pt analysis \citep[see e.g.][]{2019PhRvD.100b3541A}.  On the other hand, the KiDS+GAMA Collaboration uses estimators for the angular power spectra, which they claim to be cleaner in terms of separation of scales ($\ell$-modes) than their real-space counterparts \citep{vanUitert:2017ieu}. Also they allow for a separation of the lensing B-modes, which due to their vanishing property can be used as a consistency check. These estimators contain many terms that depend on the survey geometry and data systematic effects, while the angular power spectra used in this work are simplified neglecting many of these survey-specific terms. Apart from that, the differences to our approach lie mostly on the treatment of galaxy bias terms and the intrinsic alignment modeling.

Our goal in this work is to extend the analysis presented in \citetalias{IST:paper1} and assess the impact that the inclusion of the XC terms can have when constraining additional cosmological models, and on the understanding of systematic effects. In practice we consider several different prescriptions for the galaxy bias, photometric-redshift uncertainties and IA's, so that we can determine whether or not the impact of the XC terms on cosmological parameter inference depends on the models used.

After briefly reviewing the \Euclid survey in \Cref{sec:Euclid-survey}, we present in \Cref{sec:building-forecasts} our two probes of choice (the WL and \GCph\ angular power spectra), and the approach we adopt to forecast parameter constraints. We then present the cosmological model in \Cref{sec:cosmo}, and systematic effects in \Cref{sec:nuisance}. Finally, we present our results in \Cref{sec:results}. We conclude in \Cref{sec:conclusion}.


\section{The \Euclid survey}\label{sec:Euclid-survey}

\Euclid is an ESA M-class space mission due for launch in 2022, whose near-infrared spectrophotometric instrument \citep{NISP_paper} and visible imager \citep{VIS_paper} will carry out a spectroscopic and a photometric galaxy survey over an area $A_\textrm{survey} = 15\,000\,\deg^2$ of the extra-galactic sky \citep{2011arXiv1110.3193L}. \Euclid's main aims are to measure the geometry of the Universe and the growth of structures up to redshift $z\sim 2$ and beyond.

In this paper, we focus on the photometric observations that will be used for both a weak gravitational and a galaxy clustering survey. Given the relatively large redshift uncertainties associated with the photometric data -- compared to spectroscopy -- the analysis of the aforementioned observables will be performed via a so-called `tomographic' approach. This consists of binning galaxies according to their colour \citep{rainbow} or redshift, which results in tomographic bins that are treated as two-dimensional (projected) data sets. A spherical harmonic decomposition can be performed on the tomographically binned data to create angular (i.e.\ spherical harmonic) power spectra. In contrast, the accuracy of \Euclid's  spectroscopy will allow us to perform galaxy clustering analyses for the spectroscopic sample in three dimensions. It is important to mention that \Euclid's spectroscopy will target objects at high redshift ($0.9 < z < 1.8$, \citetalias{IST:paper1}), while photometric observations will start at much lower redshift. This motivates the consideration of the complementary information brought by photometric galaxy clustering. The \Euclid cosmological probes are therefore three: WL, \GCph, and spectroscopic galaxy clustering (\GCs).

Of particular interest is the XC between WL and \GCph, because they probe the same observed volume. In this work we focus on these XC, whilst a proper treatment of the XC between the \GCph\ and \GCs, and the XC between the \GCs  and the WL measurements, are left for future work. The modeling of WL and \GCph\ and the recipe used to compute the forecasts for \Euclid are described in next section. 


\section{Building forecasts for \Euclid}\label{sec:building-forecasts}
In this work we follow for the most part the forecasting recipe presented in great detail in \citetalias{IST:paper1}. We adopt the same Fisher matrix formalism for the computation of the forecasts, as well as the forecasting codes validated therein. However, we include in this work some important updates, which primarily concern systematic effects such as the implementation of a more realistic galaxy bias model and the inclusion of additional uncertainties on the mean redshift of the tomographic bins caused by potential errors in the photometric redshift determination. These modifications are described in detail in \Cref{sec:bias} and \Cref{sec:photoz} respectively.

For the redshift distribution of galaxies we use the same set-up as in \citetalias{IST:paper1}. Galaxies are divided into $N_z = 10$ tomographic bins as a function of redshift; each bin is equi-populated (i.e. the same total number of galaxies per bin) with respect to the true (spectroscopic) redshift:
\begin{equation}
    n^{\rm true} (z) \propto \left(\frac{z}{z_0}\right)^2\,\text{exp}\left[-\left(\frac{z}{z_0}\right)^{3/2}\right]\,,
\end{equation}
where $z_0 = 0.9/\sqrt{2}$ and the surface density of galaxies is $\bar{n}=30$ galaxies per arcmin$^2$. The true (spectroscopic) redshift distribution is then convolved with a sum of two Gaussian distributions \citepalias[see Eq. 115 in][]{IST:paper1}, to provide the observed galaxy distributions in each tomographic photo-$z$ bin, accounting for photometric-redshift errors and the fraction of outliers.

Our observables are the tomographically binned projected angular power spectrum, $C_{ij}(\ell)$, where $\ell$ is the angular multipole, and $i,j$ labels redshift pairs of tomographic bins. This formalism is the same for WL,
\GCph, and the XC terms, with the three cases differing only by the kernels used in the projection from
the power spectrum of matter perturbations, $P_{\delta\delta}$, to the spherical harmonic-space observable, as detailed in \citetalias{IST:paper1}. Under the Limber, flat-sky and spatially flat approximations \citep{2017MNRAS.469.2737K,2017MNRAS.472.2126K, 2018PhRvD..98b3522T}, these projections can be expressed as
\begin{equation}
     C^{AB}_{ij}(\ell) = \int{\rm d}z\,\frac{W^{A}_i(z)W^{B}_j(z)}{H(z)r^2(z)}P_{\delta\delta}\left[\frac{\ell+1/2}{r(z)},z\right]\,,
\end{equation}
where $A$ and $B$ stand for WL and \GCph, $r$ denotes the comoving distance, and $H$ the Hubble parameter. We also ignore reduced shear and magnification effects \citep{2019arXiv191207326D}. 

The WL power spectrum contains contributions from cosmic shear and
intrinsic galaxy alignments. We assume the latter is caused by a change in galaxy
ellipticity that is linear in the density field. Within this framework, the
density-intrinsic and intrinsic-intrinsic 3D power spectra, $P_{\delta
\textrm{I}}$ and $P_{\textrm{II}}$, respectively, are defined. These depend
linearly on the density power spectrum $P_{\delta\delta}$, with $P_{\delta
\textrm{I}} = -A(z) P_{\delta\delta}$, and $P_\textrm{II} = [-A(z)]^2 P_{\delta\delta}$.
For the redshift-dependent amplitude parameter $A(z)$, we use the model specified 
in \Cref{sec:IA}.

One of the primary sources of uncertainty for galaxy clustering is galaxy bias, i.e. the relation
between the galaxy distribution and the underlying total matter distribution. Our bias models are discussed in
\Cref{sec:bias}.

We use the same redshift bins and number density for both WL and \GCph\ 
analyses. In practice, this is an over-simplification, since lensing and
clustering will apply different probe-specific selection criteria and cuts
resulting in different samples. For the present Fisher-matrix analysis,
however, we limit ourselves to the same sample for both probes.

In the following we will use two different combinations of this observable: WL+\GCph\ where we consider the two completely independent and we simply add together the respective Fisher matrices, and WL+\GCph+XC. In the latter case, we include the XC terms i.e. we consider the full Gaussian covariance matrix, accounting for all correlations
between angular scales, redshift combinations, and correlations between the
different observables, thus also including the cross-covariance:
\begin{align}
    {\rm Cov}&\left[C_{ij}^{AB}(\ell),C_{kl}^{A'B'}(\ell')\right]=\nonumber\\
    =&\frac{\delta_{\ell\ell'}^{\rm K}}{(2\ell+1)f_{\rm sky}\Delta \ell}\left\{\left[C_{ik}^{AA'}(\ell)+N_{ik}^{AA'}(\ell)\right]\left[C_{jl}^{BB'}(\ell')+N_{jl}^{BB'}(\ell')\right]\right.\nonumber\\
    &+\left.\left[C_{il}^{AB'}(\ell)+N_{il}^{AB'}(\ell)\right]\left[C_{jk}^{BA'}(\ell')+N_{jk}^{BA'}(\ell')\right]\right\}\,,
\end{align}
where $A,B,A',B'$ stand for WL and \GCph, $i,j,k,l$ run over all tomographic bins, $\delta_{\ell\ell'}^{\rm K}$ denotes de Kronecker delta of $\ell$ and $\ell'$, $f_{\rm sky}$ represents the fraction of the sky observed by \Euclid, and $\Delta \ell$ the width of the multipole bins. The noise terms $N_{ij}^{AB}(\ell)$ take the form $\sigma_{\epsilon}^2\delta_{ij}^{\rm K}/\bar{n}_i$ for WL, where the variance of observed ellipticities is $\sigma_{\epsilon}=0.21$, and $\delta_{ij}^{\rm K}/\bar{n}_i$ for \GCph. We assume that the Poisson errors on WL and \GCph\ are uncorrelated, which yields a vanishing noise for XC. More details are laid out in \citetalias{IST:paper1}. 

In all this work we will consider two possible scenarios: an optimistic case and a pessimistic one. Following \citetalias{IST:paper1}, we define the optimistic case as the analysis including all multipoles from $\ell=10$ to $\ell=5000$ for WL and the multipoles from $\ell=10$ to $\ell=3000$ for \GCph\ and XC. In the pessimistic case we limit the maximum multipole to $1500$ for WL and $750$ for \GCph\ and XC.

It is important to mention that a joint analysis of several probes implies a large data vector, which requires a large covariance matrix. In this work we follow \citetalias{IST:paper1} in using a theoretical Gaussian covariance matrix for the observables, which makes the estimation of the covariance matrix easy. However, we must be aware that estimating the joint covariance for analyses with real measurements may become much more difficult than for single-probe analyses; in particular when the covariance is estimated from simulations. In these cases we must ensure that the constraining power brought by the different elements of the data vector is large enough to compensate for the added difficulty on the estimation of the covariance. A simple approach that could be followed is to use only the galaxies that provide the highest signal-to-noise for each probe: use only the sources located at high redshift for WL and the lenses located at low redshift (better photometric redshift estimates) for \GCph. Such selection could significantly reduce the dimensionality of the data vector and the covariance and still keep nearly the same constraining power. This was the approach used in \citet{DESY1-GCWL}. In this work we want to study the maximum impact of the XC terms. Therefore, we consider the best-case scenario where we can include all the information from all probes without considering the additional difficulty in estimating the covariance matrix. A detailed analysis will need to be done in the future to determine which fraction of the information from each probe can enter into the data vector in order to still be able to accurately estimate the covariance matrix. However, such analysis is beyond the scope of this work. 


\section{Cosmological models}\label{sec:cosmo}

We investigate the impact of the XC terms using the cosmological models discussed in \citetalias{IST:paper1}. As baseline, we consider a spatially flat Universe filled with cold dark matter and dark energy. We  approximate the dark energy equation of state parameter following the popular parameterisation \citep{Chevallier_Polarski_2001,2005PhRvD..72d3529L}
\begin{equation}\label{eq:cpl}
w(z)=w_0+w_a\frac{z}{1+z}\,.
\end{equation}
In addition to the two dark energy parameters, $w_0$ and $w_a$, the cosmological model is fully described at the background level by the total matter density at present time, $\Omegam$, 
and the dimensionless Hubble constant, $h$, for which we have $H_0 = 100\,h\,{\rm km\,s^{-1}\,Mpc^{-1}}$; where our notation is that if no time-dependence is specified for a given parameter, we consider this parameter as computed at $z=0$.

In addition, the parameters needed to describe the regime of linear perturbations are: the baryon density, $\Omegab$; the slope of the primordial spectrum, $n_{\textrm{s}}$; and the RMS of matter fluctuations on spheres of $8\,h^{-1}\,\mathrm{Mpc}$ radius, $\sigma_8$. Concerning the impact of dark energy on matter perturbations, we consider a dynamically evolving, minimally-coupled scalar field, called Quintessence, where we assume its sound speed is equal to the speed of light, and that it has vanishing anisotropic stress. This implies that we can neglect any fluctuations in the dark energy fluid in our analysis. Moreover, we allow for the equation-of-state parameter to cross $w(z)=-1$, using the prescription of \citet{Hu:2007pj}.

In addition to this baseline cosmological model, we also explore two extensions, as done in \citetalias{IST:paper1}:
\begin{itemize}
    \item non-flat models, where the curvature parameter $\Omega_{K,0}$ is nonzero. In this case we relax the flatness assumption and add the dark energy density $\Omega_{{\rm DE},0}$ as an additional free parameter;
    \item a modified gravity model with deviations in the standard growth with respect to $\Lambda$CDM. We parameterise the growth rate $f(z)$ of linear density perturbations in terms of the growth index $\gamma$ \citep{1991MNRAS.251..128L}, defined as
    \begin{equation}\label{eq:fgrowth}
        \gamma = \frac{\ln f(z)}{\ln\Omega_{\rm m}(z)} \, ,
    \end{equation}
    with $\Omega_{\rm m}(z)\equiv\Omegam(1+z)^3H_0^2/H^2(z)$.
    The growth rate $f$ is solution of the equation
\begin{equation}
f'(z)-\frac{f(z)^2}{1+z}-\left[\frac{2}{1+z}-\frac{H'(z)}{H(z)}\right]f(z)+\frac{3}{2}\frac{\Omega_{\rm m}(z)}{1+z}=0\,,
\end{equation}
where prime refers to the derivative with respect to $z$, and in general relativity we have $\gamma \approx 0.55$.
\end{itemize}

Such a modification was implemented in \citetalias{IST:paper1} via a rescaling of the power spectrum $P(k,z)$ in the following way  
    \begin{equation}\label{eq:powmg}
       P^{\rm MG}(k, z) = P(k, z) \left[\frac{D_{\rm MG}(z;\gamma)}{D(z)}\right]^2\, ,
    \end{equation}
where $D_{\rm MG}(z;\gamma)$ is the growth factor for modified gravity obtained by integrating \Cref{eq:fgrowth} for a given $\gamma$, keeping in mind that $f(z)\equiv-d\ln D(z)/d\ln(1+z)$.

The fiducial values for the cosmological parameters vector $\vec p$ also follow \citetalias{IST:paper1}, with
\begin{align}\label{eq.paramsfid}
\vec{p}&=\{\Omegam,\,\Omegab,\,w_0,\,w_a,\,h,\,n_{\rm s},\,\sigma_8,\,\Omega_{{\rm DE},0},\,\gamma\}\nonumber\\
&=\{0.32,\,0.05,\,-1.0,\,0.0,\,0.67,\,0.96,\,0.816,\,0.68,\,0.55\}\,,
\end{align}
where the last two parameters are considered only in the extended models discussed above.
In addition, we fix the sum of the neutrino masses to $\sum m_{\nu}=0.06\,$eV. It is important to note that the linear growth factor depends on both redshift and scale in the presence of massive neutrinos. However, we follow \citetalias{IST:paper1} in neglecting this small effect, given the neutrino masses considered in this analysis, and instead compute the linear growth factor in the massless limit. All fiducial values correspond to the ones provided in \citet{Ade:2015xua}.


\section{Systematic effects}\label{sec:nuisance}

While the link between the primary \Euclid probes and cosmology is well defined, we necessarily have to account for known systematic effects that can bias our results if not considered properly. In this section we focus on three systematic effects: the galaxy bias, the intrinsic alignment of galaxies, and the uncertainties in the mean of the photometric galaxy distributions in each tomographic bin. We model each of these in terms of parameterised functions where the additional parameters are known as ``nuisance parameters''. This will allow us to marginalise these effects out of our analysis.

In addition to these three mentioned effects, other sources of systematic
uncertainties could be considered, but we do not account for these in this
paper. As an example, we neglect magnification
\citep{2010A&A...523A..28K,2014PhRvD..89b3515L,2015ApJ...801...44Z,2016arXiv161110326G}
and relativistic effects
\citep{2014PhRvD..90b3513Y,2014CQGra..31w4002B,2016NatPh..12..346A,2017MNRAS.470.2822A}. Note that relativistic effects become relevant at large scales, specially for \GCph. To minimise the
impact of neglecting these effects on our results, we exclude the largest scales from
our analysis, limiting our photometric probes to $\ell\geq10$.

It is important to mention that in this work we refer to systematic effects as (astro)physical systematic uncertaintites; i.e. we only consider systematic effects originated by physical processes. In order for \Euclid to reach its expectations, we will have to overcome many observational systematic effects \citep[e.g.][]{2013MNRAS.431.3103C,2020A&A...635A.139E}, like the removal of foregrounds or the image processing. One major observational challenge for the \Euclid photometric survey will be the use of anisotropic ground-based optical data to obtain the photometric redshift estimates of the galaxies detected by the \Euclid imager. However, in this work we focus on the systematic effects with a physical origin, since they are intrinsically linked to the signal, while the analysis of observational systematic effects and how we can minimize their impact is left for future work.

\subsection{Bias modeling}
\label{sec:bias}
Weak lensing observations directly trace the underlying matter distribution $\delta_{\rm m}$, however the same does not apply for galaxy clustering. This is because galaxy clustering relies on observations of the light from galaxies, which is only a biased proxy of $\delta_{\rm m}$ \citep{1987MNRAS.227....1K}. Thus, in order to obtain theoretical predictions for galaxy clustering observations, the galaxy distribution $\delta_{\rm g}$ needs to be related to the matter distribution via a bias function, $b$. It is in general given by
\begin{equation}
    \delta_{\rm g}(z) = b(z) \delta_{\rm m}(z)\,,
\end{equation}
where we neglect the possible dependence of the bias on the scale $k$, and we assume a linear relation between the matter and galaxy distributions. Note that a linear bias approximation is sufficiently accurate for large scales \citep{DESY1-GCWL}. However, when adding very small scales into the analysis, or using for instance spectroscopic galaxy clustering, a more detailed modeling of the galaxy bias is required \citep[see e.g.][]{2017MNRAS.464.1640S}. One of the approaches to this modeling is through perturbation theory, which introduces a nonlinear and nonlocal galaxy bias (\citealp{2016JCAP...02..018L,2012PhRvD..85h3509C,2013PhRvD..87h3002S}. See also \citealp{2018PhR...733....1D} for an extensive review). 

Constraints coming from galaxy clustering alone will be affected by the marginalisation over nuisance parameter constraints used to model $b(z)$, as these will be degenerate with cosmological parameters. However the inclusion of weak lensing information (which does not depend on galaxy bias) and the XC between these two probes breaks these degeneracies and therefore improves the constraints on cosmological parameters, while also providing information on the galaxy bias itself.

In this paper we are interested in quantifying the increase in information caused by including the XC terms, and understanding if the use of this additional information will allow us to improve our knowledge on $b(z)$, within the context of different possible bias models. To achieve this purpose we investigate the impact of cross-correlations on two different parameterisations of $b(z)$:
\begin{itemize}
    \item Baseline (binned) bias \citepalias{IST:paper1}, where the bias is assumed to be constant within each of the redshift bins in true redshift, i.e.
    \begin{equation}
        b(z_i\leq z\leq z_{i+1}) = b_i\, ,
    \end{equation}
    with $z_i$ and $z_{i+1}$ the boundaries of the $i$th redshift bin in true redshift.
    \item Flagship bias, where we use a fitting function in agreement with the measurements obtained from the Flagship simulation of the \Euclid survey:\footnote{Euclid Collaboration, in preparation.}
    \begin{equation}
        b(z) = A+\frac{B}{1+\text{exp}\left[-(z-D)C\right]}\,,
    \label{eq: biasflag}
    \end{equation}
    where $A,B,C$, and $D$ are nuisance parameters.
\end{itemize}

In the first case, we choose a fiducial for our nuisance parameters corresponding to the choice
\begin{equation}
    b(z) = \sqrt{1+z}\, .
\end{equation}
Therefore, the fiducial values of the nuisance parameters are
\begin{equation}
 b_i=\sqrt{1+\bar{z}_i}\, ,
\end{equation}
with $\bar{z}_i$ the mean redshift value of each redshift bin in true redshift.

For the Flagship galaxy bias model we use fiducial parameters measured directly from the simulation, which are: $A=1.0$, $B=2.5$, $C=2.8$, and $D=1.6$. These values have been obtained by selecting all galaxies from the Flagship simulation with a magnitude in the \Euclid VIS band less than 24.5. This corresponds to the magnitude cut where extended sources will be detected at 10\,$\sigma$ in 4 exposures lasting 565 seconds each\,\citep{VIS_paper}. Once the galaxies from the simulation have been selected, we measure their galaxy clustering projected angular power spectra at different redshifts. We then obtain the galaxy bias by computing the ratio of these spectra over the theoretical matter predictions.  

It is important to note that in the Flagship case we make the assumption that we will be able to parameterise the redshift evolution of the galaxy bias. In the binned case, on the other hand, we consider several free parameters (one for each redshift bin) without attempting to model the redshift evolution within each bin. Therefore, we are not only considering two different fiducial functions for our galaxy bias evolution, but also testing the role of XC when our ability to parameterise the redshift evolution of galaxy bias is different.

\begin{figure}
\centering
    \includegraphics[width=\columnwidth]{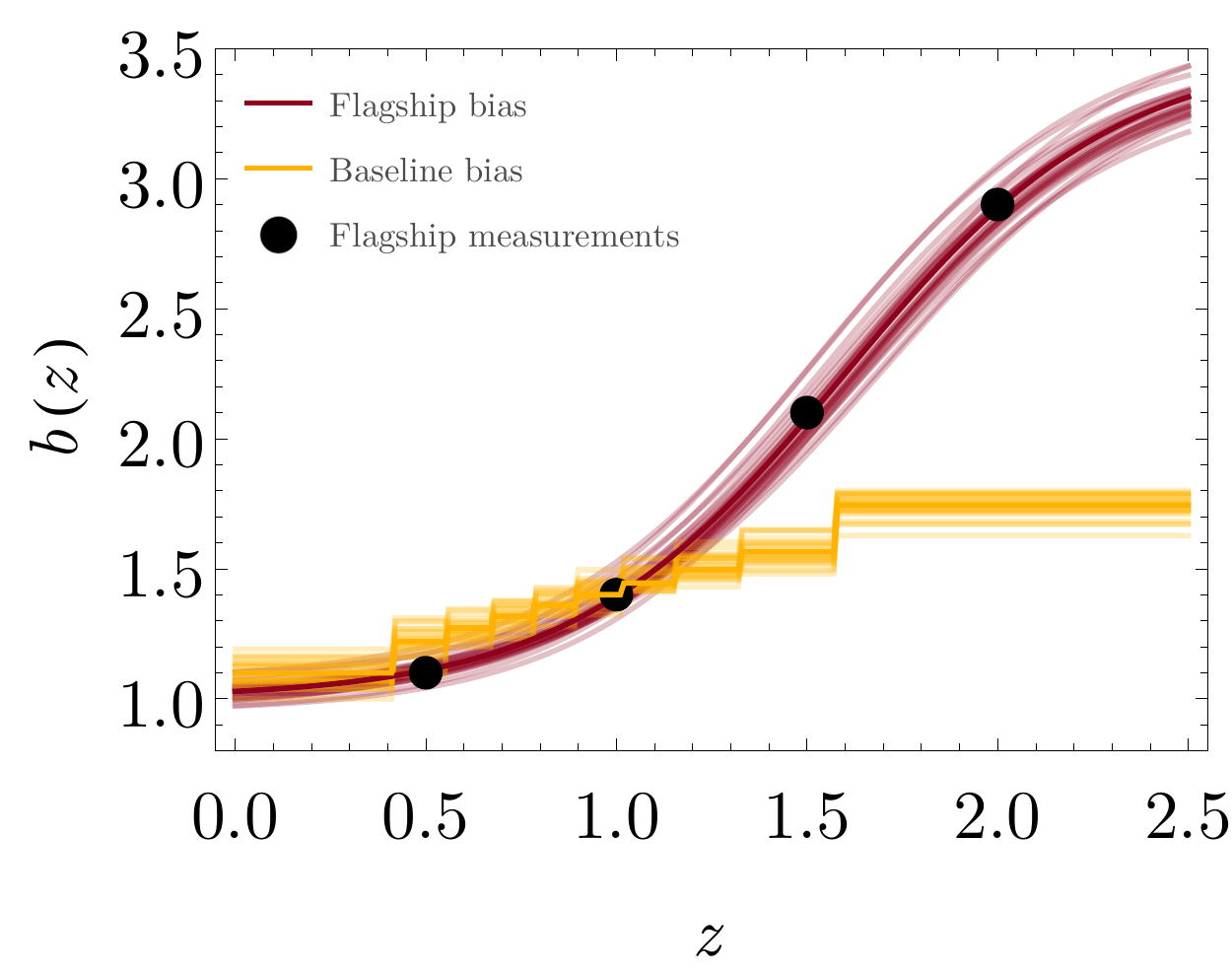}
    \caption{The plot shows the different bias parameterisations used in the paper, i.e.\ the binned bias (orange line) and the Flagship bias (red line), for the fiducial values of their respective parameters, as well as the measurements of the galaxy bias in the Flagship simulation (black dots). For illustrative purposes, further 25 lines for each case are represented alongside the fiducials, showing how the galaxy bias changes when we allow for a Gaussian dispersion of 5\% on the bias nuisance parameters.}\label{fig:biasplot}
\end{figure}

In \Cref{fig:biasplot} we show the fiducial galaxy bias for both parameterisations. For illustrative purposes we also show in \Cref{fig:biasplot} the trend in redshift of the binned and Flagship models when the bias nuisance parameters are varied, with respect to their fiducial values, with a random Gaussian dispersion of 5\%. It is important to mention that the trend of the Flagship galaxy bias beyond $z=2$ is caused by the extrapolation of the analytic parameterisation used to fit the measurements on the simulation. However, the number density of galaxies in this region is very low, which implies that the extrapolation used will have a negligible impact on the final results.

\begin{table*}
\begin{center}
\caption{Summary of nuisance parameters, together with their fiducial values, considered in this analysis.}
\resizebox{2\columnwidth}{!}{
\begin{tabular}{c|cccccccccccccc|ccc|c}
\hline
\multicolumn{1}{c}{} &\multicolumn{14}{c}{Galaxy bias} & \multicolumn{3}{c}{Intrinsic alignment} & \multicolumn{1}{c}{Photometric redshifts}\\
\hline
Parameter & $b_1$ & $b_2$ & $b_3$ & $b_4$ & $b_5$ & $b_6$ & $b_7$ & $b_8$ & $b_9$ & $b_{10}$ & $A$ & $B$ & $C$ & $D$ & $\mathcal{A}_{\rm IA}$ & $\eta_{\rm IA}$ & $\beta_{\rm IA}$ & $\Delta z_i,\,i\in[1,10]$\\
Fiducial value & 1.10 & 1.22 & 1.27 & 1.32 & 1.36 & 1.40 & 1.44 & 1.50 & 1.57 & 1.74 & 1.0 & 2.5 & 2.8 & 1.6 & 1.72 & -0.41 & 2.17 & 0.0\\
\hline
\end{tabular}
}
\label{tab: summary_params}
\end{center}
\end{table*}

\subsection{Intrinsic alignment} \label{sec:IA}

In \Cref{sec:building-forecasts}, we defined the IA amplitude parameter $A$. In our approach, this parameter takes the form given in \citetalias{IST:paper1},
\begin{equation}
A(z) = \frac{\mathcal{A}_{\rm IA} \mathcal{C}_{\rm IA} \Omegam \mathcal{F}_{\rm IA}(z)}{D(z)}\,, 
\label{eq: aiamodel}
\end{equation}
where $\mathcal{C}_{\rm IA}$ is a normalisation parameter which we set as $\mathcal{C}_{\rm IA} = 0.0134$, while $D(z)$ is the growth factor, and $\mathcal{A}_{\rm IA}$ is a nuisance parameter fixing the overall amplitude of the IA contribution. The function $\mathcal{F}_{\rm IA}$ sets the redshift dependence of the IA contribution which can be conveniently modeled as
\begin{equation}
\mathcal{F}_{\rm IA}(z) = (1 + z)^{\eta_{\rm IA}} \left [ \frac{\langle L \rangle(z)}{L_{\star}(z)}\right ]^{\beta_{\rm IA}}\,,
\label{eq: fia}
\end{equation}
with $\langle L \rangle(z)/L_{\star}(z)$ the redshift-dependent ratio between the average source luminosity and the characteristic scale of the luminosity function. \cref{eq: fia,eq: aiamodel} reduce to the nonlinear linear alignment model \citep{2007MNRAS.381.1197H,2007NJPh....9..444B} for $\eta_{\rm IA} = \beta_{\rm IA} = 0$ (i.e.\ $\mathcal{F}_{\rm IA} = 1$), while the additional scaling with $z$ and the luminosity has been introduced to improve the fit to both low-redshift data and numerical simulations \citep{iareview1}. We will refer to this model as eNLA in the following, setting the nuisance parameters to the following fiducial values
\begin{equation}
\{\mathcal{A}_{\rm IA}, \, \eta_{\rm IA}, \, \beta_{\rm IA}\} = \{1.72, \, -0.41, \, 2.17\} \, ,
\end{equation}
in accordance with the recent fit to the IA measured in the Horizon-AGN simulation \citep{2015MNRAS.454.2736C}. It is important to mention that we expect the amplitude of IA, $\mathcal{A}_{\rm IA}$, to be smaller in practice \citep[see][for a detailed discussion on the amplitude of IA for different types of galaxies]{MAFortuna}, but we keep a higher value in this analysis to study the role of XCs when IA are important.

\subsection{Photometric redshifts}\label{sec:photoz}
The accuracy of photometric redshifts is crucial to the exploitation of the galaxy clustering and weak lensing power spectra. To mitigate the effect of potential unknown biases in the photo-$z$ algorithms, we follow \citet{DESY1-WL,DESY1-GCWL} who introduced nuisance parameters for the biases/shifts of the mean redshifts of each photo-$z$ bin. That is:
\be
n_i^{\rm true}(z) =n_i(z - \Delta z_i)\,,
\ee
where $\Delta z_i$ is a nuisance parameter for each redshift bin, and $n_i^{\rm true}$ is the true galaxy distribution. This change of the galaxy redshift distribution is going to impact galaxy clustering and weak lensing predictions through their kernels. $\Delta z_i>0$ will generally lead to a higher amplitude for clustering and a lower amplitude for weak lensing, as the galaxies are shifted to lower redshifts. Because the XC terms increase the number of spectra used in a likelihood analysis with respect to the number of redshift bins (and thus the number of nuisance parameters), it may be expected that including these terms will improve the constraints on these nuisance parameters.

Before moving to the results section we summarize all the nuisance parameters considered in this analysis, together with their fiducial values, in \Cref{tab: summary_params}.



\section{Results}\label{sec:results}
In this section, we present the main results of our analysis, i.e.\ the improvement in parameter constraints using the \Euclid WL and \GCph\ probes when also including their XC terms, instead of simply considering the two probes as independent. We highlight, in particular, the importance of the XC terms in constraining the nuisance parameters since they also contain astrophysical information, for example in testing galaxy formation scenarios. In this section, unless otherwise stated, all the plots refer to the `optimistic' case.

\subsection{Baseline specifications}\label{sec:baselineres}

\begin{figure}
\centering
    \includegraphics[width=\columnwidth]{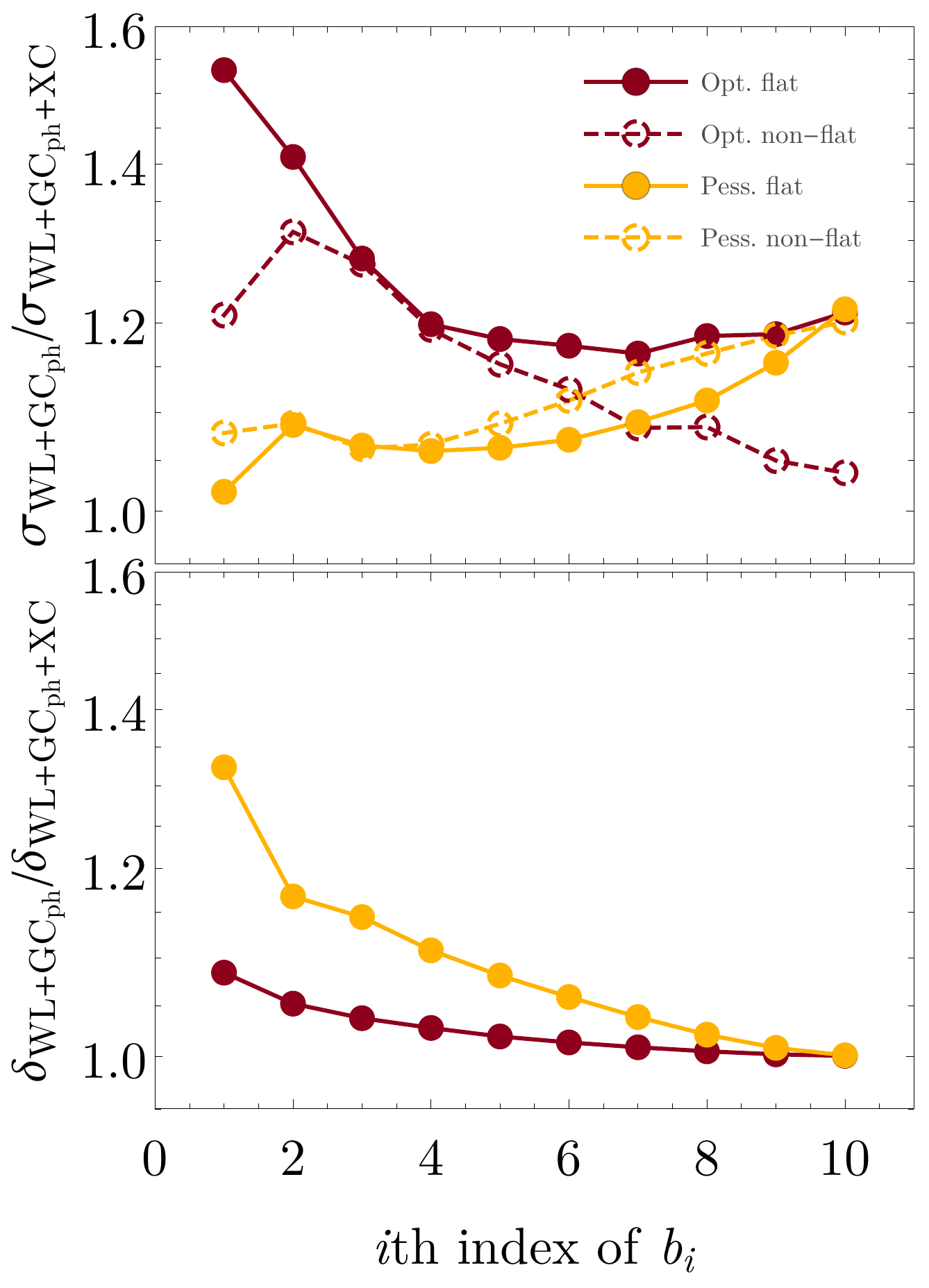}
    \caption{Ratio of the marginalised (top panel) and unmarginalised (bottom panel) forecast uncertainties on the $b_i$ bias parameters between WL+\GCph\ and WL+\GCph+XC, in the pessimistic (red lines) and optimistic (yellow lines) cases. We show in this plot results for both the flat Universe model (solid lines) and the non-flat case (dashed lines).}\label{fig:biasratio}
\end{figure}

Let us start by considering the baseline specifications described above. These are the
same as adopted in \citetalias{IST:paper1}, where the impact of XC terms on cosmological parameter constraints was discussed, finding an improvement in the FoM\,\footnote{In this work we follow the definition of FoM used in \citetalias{IST:paper1}, where it is given by $\text{FoM}=\sqrt{\det\left(\tilde{F}_{w_0w_a}\right)}$, with $\tilde{F}_{w_0w_a}$ being the marginalised Fisher submatrix for $w_0$ and $w_a$.} of a factor $5.7$ ($4.4$) for the pessimistic (optimistic) case, when a flat cosmology is assumed. Here we focus instead on the impact of the XC terms on both the cosmological and nuisance parameters.
In the top panel of \Cref{fig:biasratio} we show the ratio on the forecast uncertainties for the bias parameters with and without XC, i.e.\ $\sigma(b_i, $WL\,+\,\GCph$)/\sigma(b_i, $WL+\GCph+XC$)$. We also report the marginalised constraints on the
$10$ bias parameters in \autoref{tab:
  biasbaseflatgr}, shown in \Cref{sec:fullres}. Comparing the results with and without XC, immediately shows the power of this additional information in reducing the error on the bias parameters. On average over the $10$ parameters, we find an error reduction of $\sim 9\%$ ($\sim 25\%$) in the pessimistic (optimistic) scenario when XC is included.

It is interesting to note that there is a qualitative trend
of $\sigma(b_i, \text{WL} \,+ \, \GCph\ )$/$\sigma(b_i, \text{WL} \, + \, \GCph\ \,+ \, \text{XC})$
with the bin redshift. In particular, we find that the above ratio increases
with $z$ in the pessimistic case (going from 1.09 to 1.21), while the trend is
reversed in the optimistic case (the ratio decreasing from 1.54 to 1.20). Moreover, the improvement brought by the XC is significantly larger for the optimistic case compared to the pessimistic one. In order to understand these effects, we have investigated the ratio of the unmarginalised constraints on the galaxy bias nuisance parameters, instead of marginalising over the cosmological and IA parameters, which is shown in the bottom panel of \Cref{fig:biasratio}. In the unmarginalised case we observe that both the optimistic and pessimistic scenarios show a similar trend as a function of redshift. Moreover, the XCs are more important in the pessimistic case. This is due to the addition of more scales in the optimistic case, which helps in constraining the parameters and therefore slightly decreases the importance of the XC. Therefore, the different behaviour in the top panel of \Cref{fig:biasratio} is entirely due to the correlations between the cosmological and nuisance parameters.

It is worth investigating whether the reduction in error caused by the inclusion of the XC is model independent. To this end, we consider the case of non-flat models, i.e.\ we still leave the fiducial model unchanged, but relax the flatness assumption adding the fractional density of dark energy, $\Omega_{{\rm DE},0}$ as an additional parameter to constrain; note that we will still assume flatness in the Limber approximated power spectra \citep{2018PhRvD..98b3522T}.
We refer again to \Cref{fig:biasratio} for the ratio of the constraining power with and without XC, while we show the marginalised errors on the bias parameters for both pessimistic and optimistic assumptions in \autoref{tab: biasbasenonflatgr}, which can be found in \Cref{sec:fullres}. As expected, adding one more parameter weakens the overall constraints due to the increased volume in parameter space and degeneracies among the full set of cosmological and nuisance parameters. This increase in the marginalised errors occurs if XC is included or not. We find on average, with respect to the flat case, a $35\%$ increase of the errors for WL+\GCph\ vs. $32\%$ when XC is added in the pessimistic case, while these numbers become $7\%$ and $17\%$ in the optimistic one. When comparing WL+\GCph\ with WL+\GCph+XC constraints for non-flat models, we again find that XC reduces the marginalised errors on the bias parameters. We also find the same overall trend with redshift for the impact of XC: with the effect being smaller than the flat case for the optimistic assumption (with average improvement $15\%$ instead of $25\%$), while the opposite takes place in the pessimistic one (the average improvement being now $12\%$ instead of $9\%$). Given the different range in multipoles between the optimistic and pessimistic cases, the degeneracies introduced by the additional parameter $\Omega_{\rm DE,0}$ are lifted in the optimistic case by the use of small scales, while in the pessimistic case XC plays a more important role, thus increasing its relevance with respect to the flat cosmology. 

\begin{table*}
\begin{center}
\caption{Constraint on the IA nuisance parameters for flat and non-flat cases, for both the pessimistic and optimistic assumptions.}
\begin{tabular}{ccccc|ccc}
\hline
 \multicolumn{8}{c}{IA parameters: flat cosmology, baseline}\\
\hline
Probe & Case & $\sigma({\cal{A}}_\mathrm{IA})$ & $\sigma(\eta_{\rm IA})$ & $\sigma(\beta_{\rm IA})$ & $\sigma({\cal{A}}_\mathrm{IA})$ & $\sigma(\eta_{\rm IA})$ & $\sigma(\beta_{\rm IA})$ \\
     &  & \multicolumn{3}{c|}{pessimistic} & \multicolumn{3}{c}{optimistic} \\
\hline
 & flat & 3.85 & 2.45 & 1.03 & 3.35 & 2.13 & 0.90  \\
WL+\GCph\ & & & & & & & \\
 & non-flat & 3.86 & 2.46 & 1.04 & 3.35 & 2.13 & 0.90 \\
\hline
 & flat & 1.39 & 0.90 & 0.34 & 0.92 & 0.60 & 0.22 \\
WL+\GCph+XC & & & & & & & \\
 & non-flat & 1.40 & 0.91 & 0.34 & 0.92 & 0.60 & 0.22 \\
\hline
\end{tabular}
\label{tab: iaparbase}
\end{center}
\end{table*}

\begin{figure}
\centering
    \includegraphics[width=1\columnwidth]{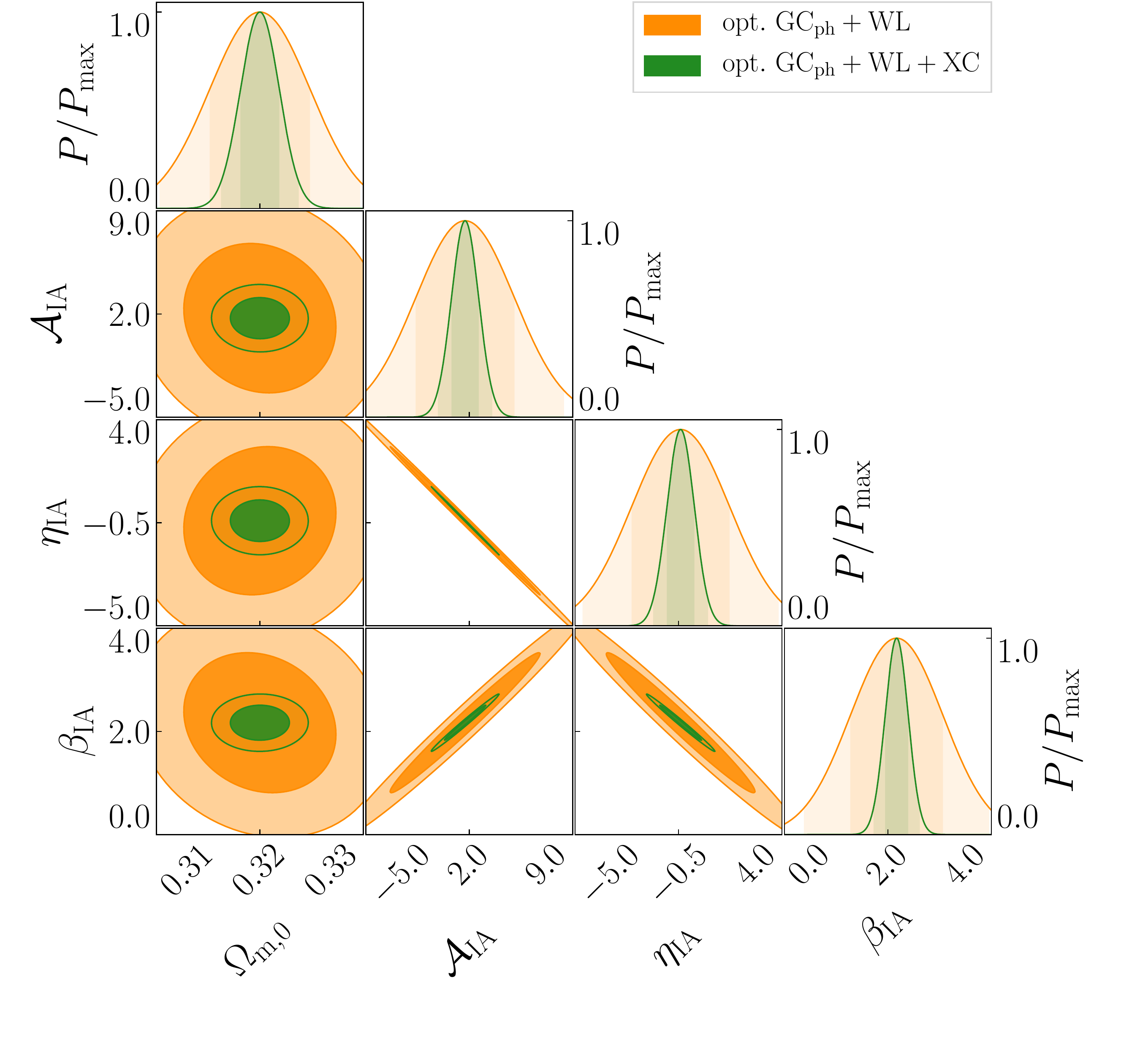}
    \caption{1$\sigma$ and 2$\sigma$ confidence contours on the optimistic, flat GR, baseline case for \GCph+WL (orange) vs. \GCph+WL+XC (green) for the 3 intrinsic alignment parameters vs. the $\Omega_{\rm m}$ parameter.
    While the IA parmeters are clearly very degenerate among themselves, $\Omega_{\rm m}$ shows little degeneracy to them, especially when XC is included.}\label{fig:IA_opt}
\end{figure}

We now discuss the constraints on the IA parameters for the eNLA model, which we summarise in \autoref{tab: iaparbase} in both the flat and non-flat cases, under both pessimistic and optimistic assumptions. We find that the marginalised errors of the IA parameters are of the same order of magnitude (if not larger) compared to the fiducial values. This is, however, not unexpected when considering that they only enter through their combination in ${\cal{A}}_{\rm IA} {\cal{F}}_{\rm IA}(z)$, so that large degeneracies are present in the amplitude terms. Indeed, we find that the correlation coefficient of, e.g.\ ${\cal{A}}_{\rm IA}$ with $(\eta_{\rm IA}, \beta_{\rm IA})$ is almost unity. In addition the \GCph\ probe is totally independent of IAs and therefore does not constrain the IA parameters at all. As a consequence, adding WL and \GCph\ has only an indirect impact on the IA constraints. To understand how this works, let us focus on the correlation between ${\cal{A}}_\mathrm{IA}$ and $\Omegam$ which is one of the largest ones in the optimistic case. When using WL alone, we find a correlation coefficient $-0.27$ which reduces to $-0.12$ when using WL+\GCph\ because of the better constraint on $\Omegam$. However, the error on ${\cal{A}}_\mathrm{IA}$ is not particularly affected by this with $\sigma({\cal{A}}_\mathrm{IA})$ reducing from $3.47$ to $3.35$, i.e.\ a $4\%$ reduction only. Instead, when XC is included, the degeneracy between $\Omegam$ and ${\cal{A}}_\mathrm{IA}$ is almost totally lifted with the correlation coefficient going down to $-0.007$ thus allowing an improvement of almost a factor of $3.5$. We show the impact of XC on IA parameter constraints in \Cref{fig:IA_opt} for the optimistic case. The inclusion of XC improves the constraints on $\{{\cal{A}}_\mathrm{IA},\,\eta_{\rm IA},\,\beta_{\rm IA}\}$, but the expected errors are smaller than the corresponding fiducial value only in the optimistic scenario. It is also worth noting that relaxing the flatness assumption does not degrade the constraints on the IA parameters. This is just a consequence of the IA parameters being almost uncorrelated with $\Omega_{{\rm DE},0}$ so that there are no further degeneracies introduced. For this same reason, the impact of XC works in the same way as the flat case, since the same qualitative argument still holds in the non-flat case.

To conclude this section, which focuses on the specifications and models under investigation in \citetalias{IST:paper1}, we want to quantify the impact of the XC terms on modified gravity constraints. We use here the phenomenological approach described in \Cref{sec:cosmo}, i.e. we consider the growth index $\gamma$. In the $\Lambda$CDM concordance model, $\gamma \approx 0.55$ with gravity described by general relativity. A deviation from this value could be indicative of phenomena associated with modified gravity. To forecast how accurately \Euclid photometric probes can constrain this parameter, we add $\gamma$ as a new parameter in the Fisher matrix by the simple extension of the general relativity recipe from \Cref{sec:cosmo}. In this case, we find a significant weakening of the constraints as a consequence of including this additional parameter, as expected. More interestingly, we also find that XC is highly efficient in improving the constraints on each cosmological parameter compared with WL+\GCph\ only. This is consistent with the findings of \citetalias{IST:paper1} for the general relativity model considered, and we now extend this result to the modified gravity case. In particular, the error on $\gamma$ reduces by a factor 1.5 in the pessimistic case showing that XC helps to constrain deviations from general relativity. The impact is less pronounced when one includes larger $\ell$, with XC reducing the error in the optimistic case by only $10\%$, due to the fact that, in the optimistic case, the constraints are now driven by the nonlinear part of the power spectrum. In this work we model the nonlinearities following the same prescriptions of \citetalias{IST:paper1}, which are obtained in the $\Lambda$CDM regime. However, an adapted recipe for nonlinearities should be applied for each modified gravity theoretical model, which could change the quantitative impact of XC depending on the modified gravity model considered. 

\begin{figure}
\centering
   \includegraphics[width=1\columnwidth]{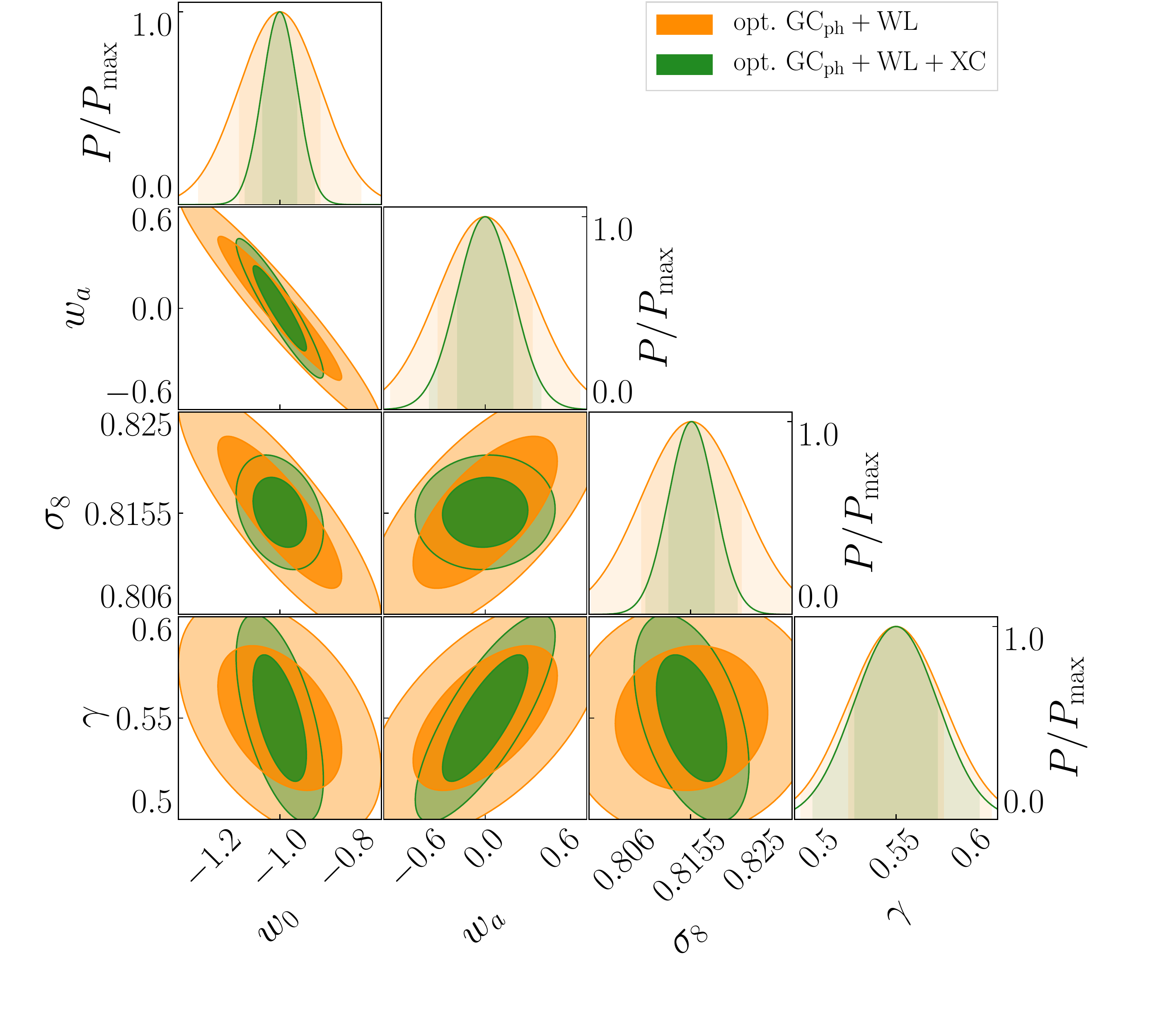}
    \caption{
    1$\sigma$ and 2$\sigma$ confidence contours on the optimistic, flat modified gravity, baseline case for \GCph+WL (orange) vs. \GCph+WL+XC (green). The modified gravity parameter $\gamma$ is not significantly better constrained when including XC, but its inclusion helps to break some degeneracies, especially with $w_0$, $w_{a}$, and $\sigma_8$.}\label{fig:triplot-gamma-w0wa}
\end{figure}

We show the effect of XC on $\gamma$ constraints in \Cref{fig:triplot-gamma-w0wa}. We report the $68\%$ errors on this and the other cosmological parameters in \autoref{tab: cosmogammaflat}, while we show the constraints on the bias parameters in \autoref{tab: biasgammaflat}. 

In contrast with the cosmological parameters, the marginalised errors on the IA parameters are not affected by the presence of the $\gamma$ parameter, and therefore the effect of XC on these is unchanged. This is a consequence of the interplay among the degeneracy of IA and cosmological parameters which is now changed with respect to the general relativity case. As a result, the weakening of the constraints on cosmological parameters does not lead to a corresponding increase in the errors for the IA parameters. 

Finally, we have also investigated the case when the flatness assumption is relaxed, leaving $\Omega_{{\rm DE},0}$ free to vary. We do not find any remarkable difference on the impact of XC, apart from the expected degradation of the constraints due to having one more parameter.

\subsection{Dependence on the galaxy bias model}
Galaxy bias enters both the \GCph\ and XC terms, therefore it is worth considering how the cosmological constraints depend upon the bias model. In the baseline specifications, $b(z)$ was modeled as a piecewise constant function with independent amplitudes in the 10 redshift bins. However, $N$-body simulations coupled with reliable models for galaxy distributions and halo occuptation statistics can provide a physically-motivated prior on the redshift dependence of the bias function. By using this information, we model $b(z)$ using \Cref{eq: biasflag}, with the four parameters $\{A,\,B,\,C,\,D\}$ as the new nuisance parameters free to vary in our Fisher analysis.

Since we reduce the number of nuisance bias parameters from 10 to 4, or in other words we assume we can parameterise the redshift evolution of the galaxy bias even within each redshift bin, we expect an improvement of the constraints on the cosmological parameters. In \autoref{tab: cosmoparflagflatgr} we report the marginalised errors for the flat general relativity case\footnote{We will discuss here only the results for the flat model since both the effects of changing the bias and the impact of XC is qualitatively the same for flat and non-flat models.} which can be compared with the corresponding table of \citetalias{IST:paper1}. Averaging over the full set of parameters, we indeed find a $17\%$ ($14\%$) reduction of the WL+\GCph\ (WL+\GCph+XC) errors for the pessimistic case, due to the more rigid bias modelling. This factor reduces to $4\%$ ($6\%$) in the optimistic case. We also find that reducing the number of bias parameters reduces the correlation between parameters, as can be appreciated by looking at the FoM. The WL+\GCph\ FoM indeed increases by $37\%$ ($15\%$) for the pessimistic (optimistic) case, while the WL+\GCph+XC FoM improves by $54\%$ ($30\%)$. Although not unexpected, this significant boost of the FoM points at the importance of constraining the galaxy bias in the \GCph\ sample.

Regardless of the bias model used, adding XC to WL and \GCph\ still stands out as the most efficient way to strengthen the constraints on the cosmological parameters, and hence also increase the FoM. In detail, the impact of XC for the pessimistic case is large, with ${\rm FoM}({\rm WL}+{\rm \GCph\!}+{\rm XC}) / {\rm FoM}({\rm WL}+{\rm \GCph}) = 5.7$ for the binned bias modeling. This factor reduces somewhat to ${\rm FoM}({\rm WL}+{\rm \GCph\!}+{\rm XC}) / {\rm FoM}({\rm WL}+{\rm \GCph}) = 4.8$ when the Flagship bias modeling is adopted. For the optimistic case, these ratios are respectively 4.4 and 3.7. This reduction of the impact of XC when moving from binned bias to Flagship bias is related to
the Flagship bias already removing part of the degeneracy between cosmological and nuisance quantities, thanks to its smaller number of parameters. The numbers are nevertheless still large enough that the impact of XC is of paramount importance. 

\begin{table*}
\centering
\caption{$68\%$ errors, without and with the XC contribution, on the nuisance parameters (both bias and IA) when the Flagship bias model is considered. We consider a flat general relativity model, and set $\{\ell_{\rm max}({\rm WL}), \ell_{\rm max}($\GCph$)\}$ according to the pessimistic and optimistic assumptions.}
\begin{tabular}{ccccccccc}
\hline
 \multicolumn{9}{c}{Nuisance parameters: flat cosmology, Flagship bias}\\
\hline
Probe & Case & $\sigma(A)$ & $\sigma(B)$ & $\sigma(C)$ & $\sigma(D)$ & $\sigma({\cal{A}}_\mathrm{IA})$ & $\sigma(\eta_{\rm IA})$ & $\sigma(\beta_{\rm IA})$ \\
\hline
 & pessimistic & 0.0086 & 0.0710 & 0.0546 & 0.0133 & 3.83 & 2.43 & 1.02 \\
WL+\GCph\ & & & & & & & & \\
 & optimistic & 0.0045 & 0.0341 & 0.0257 & 0.0053 & 3.34 & 2.13 & 0.90 \\
\hline
 & pessimistic & 0.0081 & 0.0618 & 0.0505 & 0.0128 & 1.38 & 0.89 & 0.34 \\
WL+\GCph+XC & & & & & & & & \\
 & optimistic & 0.0028 & 0.0280 & 0.0214 & 0.0049 & 0.91 & 0.59 & 0.22 \\
\hline
\end{tabular}
\label{tab: nuisparflagflatgr}
\end{table*}

Let us now discuss the constraints on the bias and IA nuisance parameters reported in \autoref{tab: nuisparflagflatgr}. Concerning the bias parameters, a direct comparison with the baseline case of \autoref{tab: biasbaseflatgr} is not possible as we use different parameterisations. We can nevertheless note that the impact of XC on the constraints is comparable to the binned bias case. Indeed, we find that adding XC reduces the errors on average by $8\%$ ($27\%$) in the pessimistic (optimistic) case which are roughly the same as what we found before. Again, this is consistent with the expectation since the effect of XC is to reduce the correlation among the bias and the cosmology, and this happens no matter which bias model is used.

Moreover, we find that the constraints on the IA nuisance parameters are not modified with respect to the binned bias case. This happens because bias and IA are two different phenomena affecting only \GCph\ and WL respectively, but not both of them. Therefore, the bias model has no impact on IA constraints and the effect of XC on those is similar to what was found in the baseline case.

In addition, we investigate the effect of the change in the bias modeling on constraints on the modified gravity parameter $\gamma$. We find $\sigma(\gamma) = 0.046$ $(0.017)$ for the pessimistic (optimistic) case using only WL\,+\,\GCph, while adding XC reduces the error to $\sigma(\gamma) = 0.036$ $(0.014)$, i.e.\ a $27\%$ ($21\%)$ improvement. XC has a different impact on $\gamma$ with respect to the binned bias case, where the improvement was of $50\%$ in the pessimistic case and of $10\%$ in the optimistic one. 
The different relevance of XC in this case is connected to the significant improvement that a change in the bias modeling brings on  $\sigma(\gamma)$. Replacing the binned bias with the Flagship one improves the constraints on $\gamma$ by a factor $\sim 1.5$ for both the pessimistic and optimistic cases when WL+\GCph+XC is used. We can therefore conclude that a reliable modelling of the galaxy bias provides a valuable help to discriminate general relativity and modified gravity models based on the growth of structures.

\subsection{Shift in cosmological parameter best fit from neglecting IA}

In \Cref{sec:IA} we described how we include the IA contribution in the theoretical predictions for cosmic shear, while in \Cref{sec:baselineres} we estimated the improvement brought by XC on the constraints of the parameters modeling this effect. Given the constraining power brought by \GCph\ and XC in addition to the WL probe, it is worth exploring if such a combination allows one to distinguish different assumptions pertaining to this contribution.

To estimate whether one can distinguish between different models of a physical mechanism the shift in the best fit of the cosmological parameters with the `wrong' model can be tested. In an MCMC (of nested sampling) analysis this could be estimated by generating mock data with a given fiducial cosmology and fitting these with theoretical predictions from a different one; the cosmological parameters will be shifted from the assumed fiducial values in order to compensate the different effects of the two cosmologies.

However, this investigation can also be performed within the Fisher matrix formalism if we deal with `nested models', where the parameter space of one of the two cosmologies is contained within that of the other \citep{Heavens:2007ka}. In this framework, the former cosmology is described by a set of parameters $\{\theta_\alpha\}$, while the latter by $\{\theta_\alpha\}\cup\{\psi_a\}$, and note that in this treatment we label the nested model parameters by indexes $\alpha,\beta,\ldots$ and the extra parameters by indexes $a,b,\ldots$ An interesting question then is: what happens to the best-fit estimates of the parameter set $\vec\theta$ if we do not properly model $\vec\psi$ in our analysis? For instance, if reality is described by the parameter $\psi_a^{\rm true}$ and we wrongly assume $\psi_a^{\rm fid}$ as fiducial cosmology, this will imply a shift in the $\vec\theta$ parameters, due to a compensation that has to account for $\vec\psi$ being kept fixed to an incorrect value. In a Fisher matrix analysis, such a shift $\delta$ on a parameter $\theta_\alpha$ can be computed via \citep[see][Appendix~A]{Camera:2016owj}
\begin{equation}\label{eq:fishift}
    \delta(\theta_\alpha)=\left(\tens F^{-1}\right)_{\alpha\beta}F_{\beta a}\left(\psi_a^{\rm fid}-\psi_a^{\rm true}\right)\, ,
\end{equation}
where it is worth emphasising that $\tens F$ is the \textit{full} Fisher matrix, containing both parameter sets $\vec\theta$ and $\vec\psi$, whereas $F_{\beta a}$ are the elements of the rectangular sub-matrix mixing $\vec\theta$ and $\vec\psi$ parameters, and summation over equal indexes is assumed.

\begin{figure*}
\centering
\includegraphics[width=\textwidth]{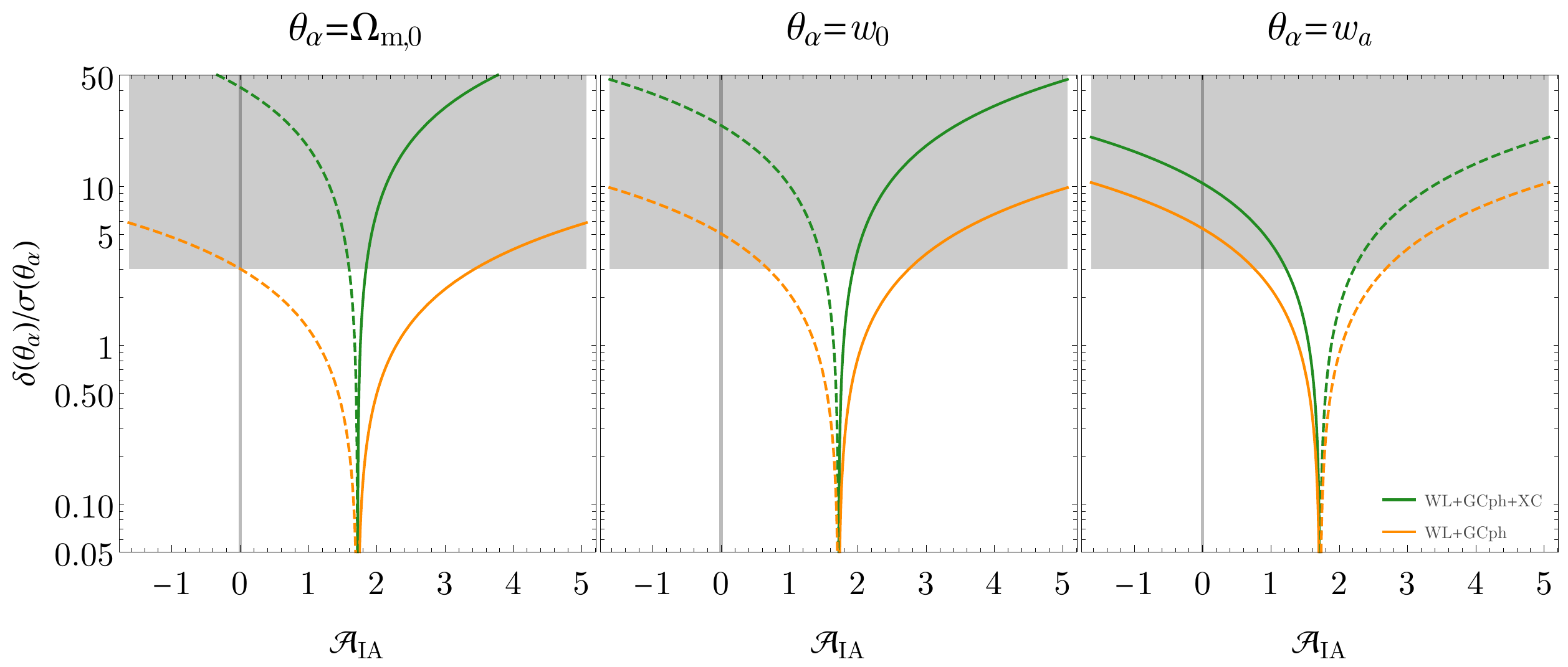}
\caption{Shifts in units of standard deviations for $\Omegam$ (left panel), $w_0$ (middle panel), and $w_a$ (right panel) due to an incorrect assumption on the IA amplitude. Results are shown for the optimistic case with WL+\GCph\ (orange) vs. WL+\GCph+XC (green). Solid (dashed) lines represent positive (negative) shifts with respect to the fiducial. The vertical gray line shows the case in which we assume no contribution from IA ($\mathcal{A}_{\rm IA}=0$). The gray-shaded region denotes the shifts larger than $3\sigma$, for which the Gaussian approximation breaks down and the corresponding shifts should be interpreted with caution (see the text for details).}
\label{fig:fishift}
\end{figure*}

We apply such methodology to investigate if the combination of \GCph, WL and their
XC is able to distinguish different amplitudes
$\mathcal{A}_{\rm IA}$ of the IA contribution. To this end, we keep the fiducial of $\beta_{\rm IA}$ and $\eta_{\rm IA}$ to our baseline, but we use
\Cref{eq:fishift} to compute the shift on the cosmological parameters caused by a wrong assumption of $\mathcal{A}_{\rm IA}$. The results for $\Omegam$, $w_0$ and $w_a$ are shown in \Cref{fig:fishift}, where we highlight the significance of this shift when
$\mathcal{A}_{\rm IA}$ is changed from the baseline fiducial.\footnote{We do not show the results for the other parameters because there are no significant differences.} We see that completely neglecting the IA contribution ($\mathcal{A}_{\rm IA}=0$)
leads to shifts of $\approx40\sigma$ on $\Omegam$, $\approx20\sigma$ on $w_0$ and $\approx10\sigma$ on $w_a$ when the full combination WL+\GCph+XC is
considered. If we do not include XC, such shifts reduce to $\approx 3\sigma$, $\approx5\sigma$, $\approx5\sigma$ respectively, thereby demonstrating how the
addition of XC is relevant not only to improve the constraining
power of the survey, but also to put to test the assumptions made in the
modelling of nuisance effects. It is important to point out here that the huge
values of the shifts on the parameters when XC is included should not be taken
literally; when the shift with respect to the fiducial values becomes large,
the Gaussian approximation on which the Fisher matrix analysis relies breaks
down. Therefore, the Fisher approximation can no longer capture the true shape of the likelihood. We consider the value of $3\sigma$ as a safe threshold, meaning that any shift beyond this limit should be interpreted just as a very large shift. We represent this region in \Cref{fig:fishift} with a gray-shaded area.

It is worth mentioning before finishing this subsection that we have also considered to apply this extended Fisher formalism to quantify the shift in cosmological parameter best fit from assuming a `wrong' fiducial galaxy bias. In more detail, we have first assumed that our galaxy bias could be modeled by a piecewise constant function.\footnote{Note that the extended Fisher formalism used in this work can only accommodate for nested models. Because of this we cannot compute the shift when changing the fiducial and the parameterisation of the galaxy bias at the same time, as it was done in \Cref{sec:bias}.} We have then used \Cref{eq:fishift} to compute the shift in cosmological parameters when we consider the fiducial $b(z)=\sqrt{1+z}$ but the `truth' is given by the Flagship fiducial described in \Cref{sec:bias}. We have obtained even larger shifts than for the IA case, which go from $4\sigma$ for $n_{\rm s}$ up to very biased (more than $40\sigma$) for $w_0$ and $w_a$ when we consider the optimistic WL+\GCph+XC case. Since XC partially removes the degeneracy between the cosmological and galaxy bias parameters, the shifts become even larger when XC is not included. It is important to recall that given the large shifts these values should only be interpreted qualitatively, since they are significantly beyond the $3\sigma$ safe threshold. These results show that our knowledge on the galaxy bias has a significant impact on the cosmological conclusions derived from the observations.

\subsection{Impact of XC on photo-z self-calibration}

In previous Sections we have implicitly assumed that photometric redshifts have been measured with perfect average accuracy, i.e. the mean true redshift of a bin is indeed equal to the mean measured redshift. However, as pointed out in \Cref{sec:photoz}, we can consider the possibility of an error $\Delta z_i$ systematically shifting the redshift of all the sources in the $i$th bin. Allowing for an arbitrary deviation, we can include 10 additional nuisance parameters $\Delta z_i$. We can then investigate both how well these quantities must be known not to degrade the FoM, and which constraints can be put on them by the XC terms.

To this end, we have recomputed the Fisher matrices for the baseline case of general relativity, adding the 10 $\Delta z_i$ nuisance parameters and fixing their fiducial values to zero.\footnote{We do not expect the results to qualitatively change for other assumptions on either the cosmological model or the galaxy bias.} We add the same Gaussian prior on each one of these nuisance parameters, and we compute the FoM as a function of the width of the prior. We finally compare the output to the case when all $\Delta z_i$ are considered known (${\rm FoM_{ref}}$, equivalent to a Dirac delta prior around the fiducial value).

\begin{figure}
\centering
    \includegraphics[width=\columnwidth]{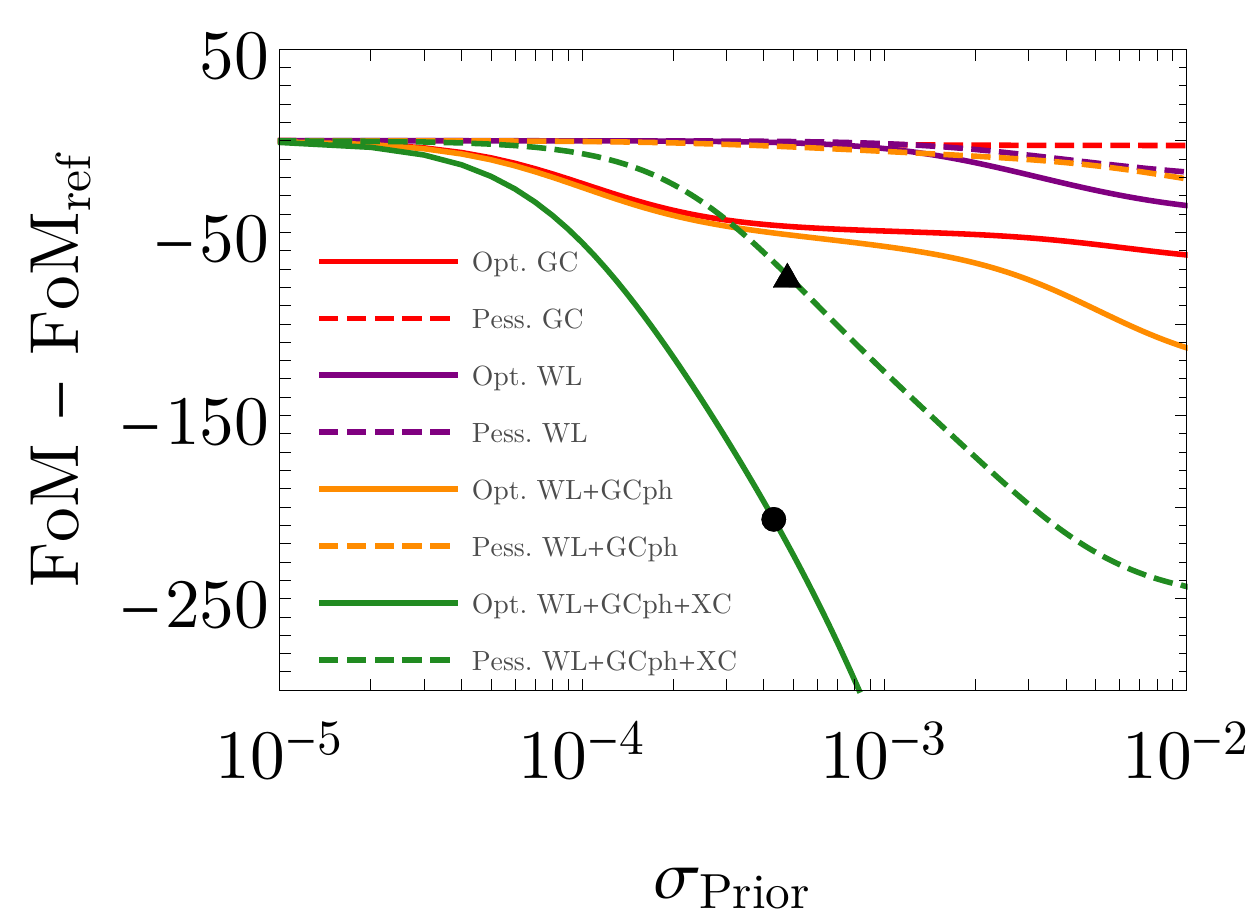}
    \caption{Absolute difference of the FoM with respect to the reference one obtained in the case when all shifts in the mean of the photometric-redshift distributions are perfectly known and equal to zero ($\Delta z_i=\sigma(\Delta z_i)=0$), for a changing value of the prior added. The results refer to \GCph\ (red), WL (purple), WL+\GCph\ (orange), and WL+\GCph+XC (green), under pessimistic (dashed) and optimistic (solid) assumptions. The black dot denotes the prior threshold for which the final FoM of the full WL+\GCph+XC combination in the optimistic case is degraded by 20\%. The black triangle represents the same threshold in the pessimistic case (see the text for details). Note that the reference FoM is larger for the optimistic case in comparison to the pessimistic case, and it increases when the XC terms are included. However, the different lines are normalised to their corresponding reference FoMs for illustrative purposes. }\label{fig:fomrationz}
\end{figure}

\autoref{fig:fomrationz} shows the absolute difference between the ${\rm FoM}$ and the corresponding reference ${\rm FoM_{ref}}$ as a function of the width of the Gaussian prior added, $\sigma_{\rm Prior}$. It provides the absolute ${\rm FoM}$ degradation for different combinations of cosmological probes both in the optimistic and pessimistic scenarios. We can observe that a strong prior is needed if one does not want to degrade the FoM by a large amount. Let us first consider the optimistic case. Starting with the full combination of WL+\GCph +XC, we require a prior on the mean of the photometric galaxy distributions smaller than $0.43\times 10^{-3}$ in order not to degrade the final FoM of 1034 by more than 20\%. This threshold is represented in \Cref{fig:fomrationz} with a black dot. Note that the prior requirement coming from the other probes may be more stringent if we require a degradation smaller than 20\% with respect to the corresponding reference FoM; for instance, we need a prior smaller than $0.31\times 10^{-3}$ for the WL+\GCph\ combination if we consider the reference WL+\GCph\ FoM, while we only need a prior smaller than $ 1.60\times 10^{-3}$ for WL alone \citep[note that this value is similar to the one provided in][and the small discrepancies might be due to the different IA modeling and forecasting recipe]{Kitching2008}. However, the combination driving the requirement on the prior is the full combination of WL+\GCph +XC, since it is the one providing the highest FoM. 

Focussing now on the pessimistic case, we require a prior smaller than $0.48\times 10^{-3}$ in order not to degrade the final WL+\GCph +XC FoM of 367 by more than 20\%. It is represented with a black triangle in \Cref{fig:fomrationz}. 

It is also important to compare the degradation of the FoM as a function of the prior for the different combinations of probes. We can see in \Cref{fig:fomrationz} that for the optimistic case the degradation appears earlier (we need a smaller prior) in the full combination of WL+\GCph +XC than in the combination WL+\GCph. At its turn, WL+\GCph\ degrades earlier than \GCph\ alone, which degrades earlier than WL. This is consistent with the values of the FoM for the different combinations of probes, since \GCph\ provides a larger FoM than WL alone, and WL+\GCph\ provides a larger FoM than \GCph\ but smaller than the full combination. Concerning the pessimistic case, we can now observe that we need a more stringent prior for WL than for \GCph, which is consistent with the fact that in the pessimistic case the FoM of WL is much larger than the \GCph\ one.

\begin{figure}
\centering
    \includegraphics[width=\columnwidth]{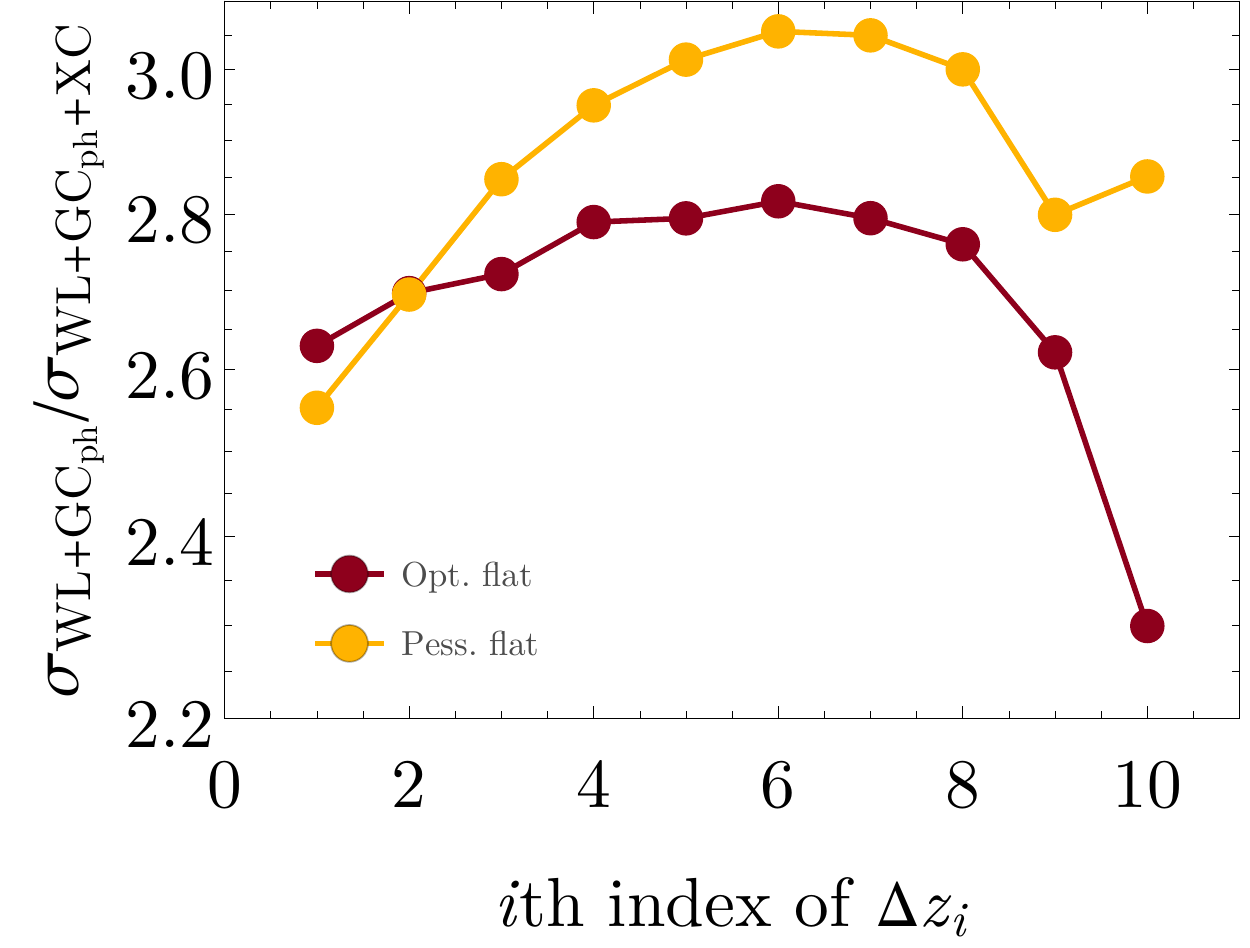}
    \caption{Ratio of the errors on $\Delta z_i$ without and with the inclusion of XC. Yellow and red lines refer to the pessimistic and optimistic scenario.}\label{fig:errrationz}
\end{figure}

Although reducing $\Delta z_i$ will be achieved by improving photometric redshifts \citep[see also][for a detailed analysis on how clustering information could help in better determining $\Delta z_i$]{Gatti2018}, it is nevertheless worth wondering whether one can use the data itself to self-calibrate/constrain $\Delta z_i$. From this point of view, it is interesting to look at how the constraints change when XC is added to WL\,+\,\GCph. The result is shown in \Cref{fig:errrationz} for both pessimistic and optimistic scenarios. In this case XC is indeed of great help reducing the error on $\Delta z_i$ by a factor $2.2 - 3.1$ as a consequence of both the increased number of observables and the information carried by the correlation among different bins. It is worth noting that the improvement of constraints due to XC is smaller for the optimistic scenario. This is due to the information brought by the additional multipoles included with respect to the pessimistic case, which already add information to constrain these nuisance parameters and reduce the impact of XC.


\section{Conclusions}\label{sec:conclusion}
In this paper, we have extensively scrutinised the impact on both cosmological and nuisance parameter estimation of the cross-correlations (XC) between two probes of the \Euclid satellite mission: weak lensing cosmic shear (WL) and the clustering of the photometric galaxy sample (\GCph).

Let us first emphasise that the XC terms have necessarily to be included in the data vector, because both WL and \GCph\ trace the same underlying cosmic structure and are, therefore, not independent from one another. This implies that a failure to include the XC terms will lead to an incorrect estimation of the constraining power, with associated consequences for model testing. The scope of this paper is to assess the impact of such XC.

In \citetalias{IST:paper1} it has been demonstrated that rather than being a nuisance, XC encodes valuable cosmological information. In that paper we found that the best cosmological constraints are obtained when XC is taken into account, leading to an increase of the FoM by more than a factor 3 for a flat Universe. Here we have focused on showing that XC is crucial also to constrain nuisance parameters.

Our main findings can be summarised as follows:
\begin{itemize}
    \item On average, the uncertainty on the galaxy bias amplitude nuisance parameters is reduced when XC is included by $\sim 9\%$ or $\sim 25\%$, for the pessimistic or optimistic scenario, respectively.
    \item The inclusion of XC makes IA parameters detectable (in the optimistic scenario), and this result is not affected by the assumptions of flatness and the validity of general relativity.
    \item A different bias model does not directly impact the effect of XC on cosmological and nuisance parameters. However, the lower number of parameters needed to describe the galaxy bias significantly affects the constraints on deviations from general relativistic growth of structures, parameterised by $\gamma$. This in turns also changes the impact of XC on this same parameter.
    \item Given the tighter constraints allowed by XC, the interplay between different parameters becomes more important, and wrong assumptions on systematic-effect parameters such as IA may lead to significant degradation of the survey \textit{accuracy} in cosmological parameter estimation.
    \item The addition of XC significantly helps in constraining the mean of the photometric-redshift distributions. However, the requirement on the knowledge of the mean is much more stringent than for WL alone in order not to degrade the final FoM.
\end{itemize}

We can conclude that the addition of XC between cosmic shear and galaxy clustering for the photometric \Euclid survey is necessary for the analyses of the future data. Not only does XC improve our knowledge of the cosmological model, but they also provide information about galaxy bias, IAs, and help in self-calibrating the photometric galaxy distributions.

\begin{acknowledgements}
This work started in the cross-correlation group, led by A.~Blanchard, M.~Kunz, and F.~Lacasa, of the Inter-Science Working Group Task-force for forecasting of the Euclid Consortium. I.~Tutusaus acknowledges support from the Spanish Ministry of Science, Innovation and Universities through grant ESP2017-89838-C3-1-R, and the H2020 programme of the European Commission through grant 776247. M.~Martinelli has received the support of a fellowship from `la Caixa' Foundation (ID 100010434), with fellowship code LCF/BQ/PI19/11690015, and the support of the Spanish Agencia Estatal de Investigacion through the grant `IFT Centro de Excelencia Severo Ochoa SEV-2016-0597'. S.~Camera acknowledges support from the `Departments of Excellence 2018-2022' Grant awarded by the Italian Ministry of Education, University and Research (\textsc{miur}) L.~232/2016. S.~Camera is supported by \textsc{miur} through Rita Levi Montalcini project `\textsc{prometheus} -- Probing and Relating Observables with Multi-wavelength Experiments To Help Enlightening the Universe's Structure'. F.~Lacasa acknowledges support by the Swiss National Science Foundation. S.~Ili\'c acknowledges financial support from the European Research Council under the European Union's Seventh Framework Programme (FP7/2007--2013)/ERC Grant Agreement No. 617656 ``Theories and Models of the Dark Sector: Dark Matter, Dark Energy and Gravity''. V.~Yankelevich acknowledges funding from the European Research Council (ERC) under the European Union’s Horizon 2020 research and innovation programme (grant agreement No.\ 769130). \AckEC
\end{acknowledgements}


\appendix


\section{Complementary results}\label{sec:fullres}

In this Appendix we show additional results on which we only comment in \autoref{sec:results}. In \autoref{tab: biasbaseflatgr} and \autoref{tab: biasbasenonflatgr} we show the $68\%$ error on the galaxy bias parameters in our baseline modeling (binned bias), for the flat and non-flat cases, respectively. \autoref{tab: cosmogammaflat} and \autoref{tab: biasgammaflat} show instead the constraints on cosmological and galaxy bias parameters, respectively, in the binned bias case, when we allow for deviations from general relativity in a flat Universe. We do not report results on the IA parameters as these are unchanged with respect to the general relativity case of \autoref{tab: iaparbase}. 

\begin{figure*}
\begin{center}
\includegraphics[width=1.3\columnwidth]{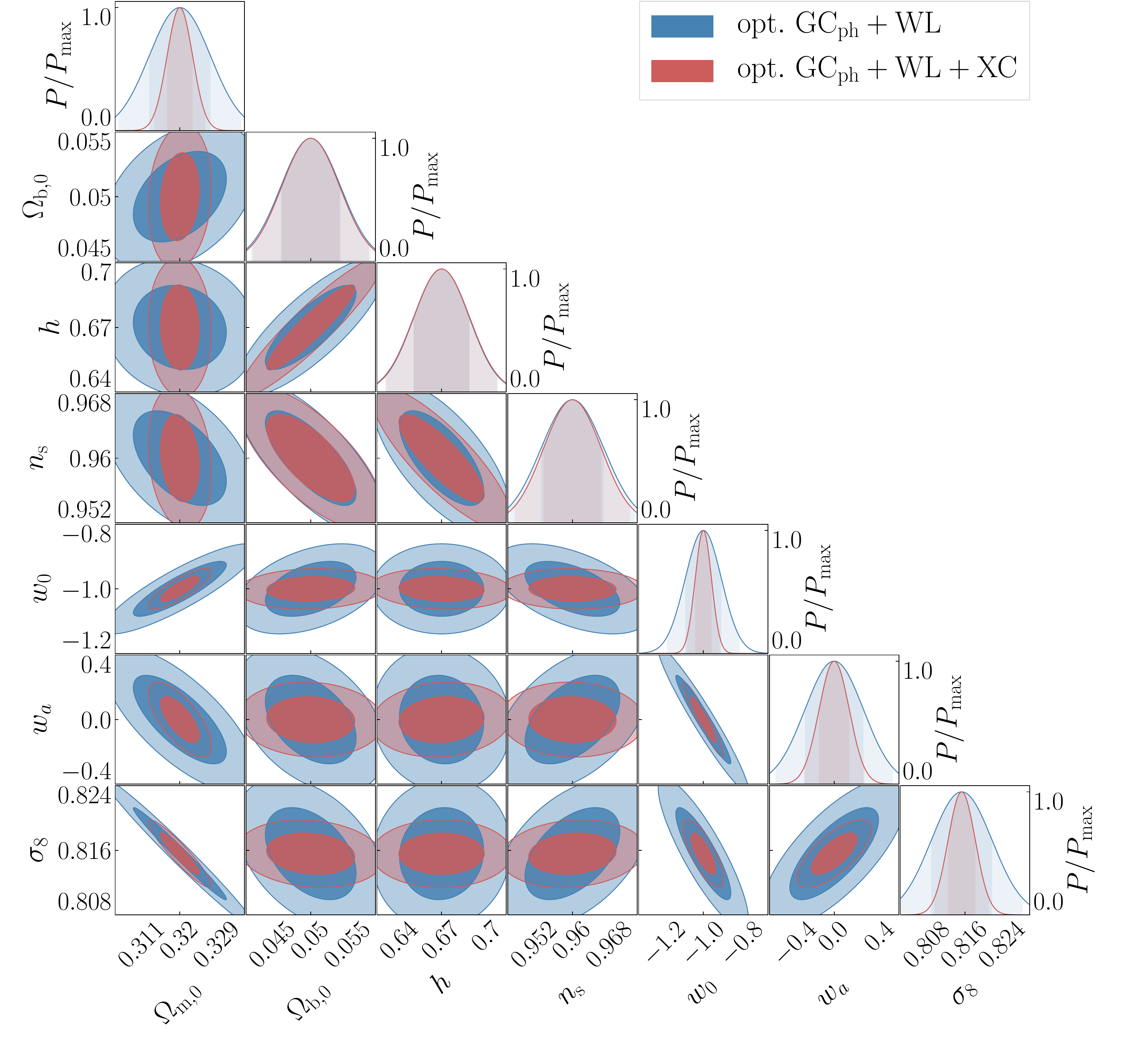}\\
\vspace{25pt}
\includegraphics[width=1.3\columnwidth]{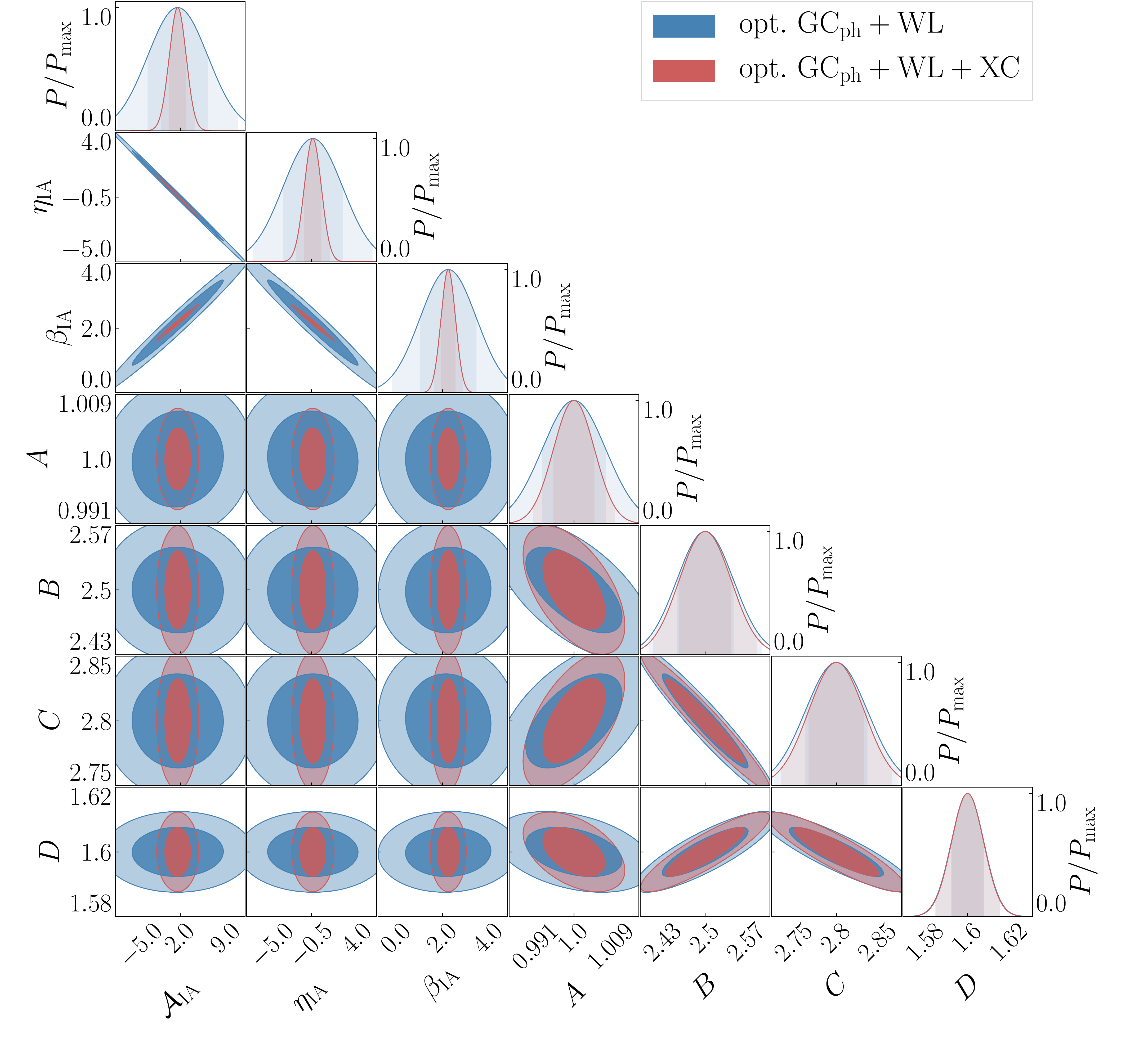}\\
\caption{{\it Top panel:} 1$\sigma$ and 2$\sigma$ confidence contours on the optimistic, flat GR case for \GCph+WL (blue) vs. \GCph+WL+XC (red) for the 7 cosmological parameters, using the Flagship bias as a fiducial galaxy bias model. {\it Bottom panel:} 1$\sigma$ and 2$\sigma$ confidence contours on the optimistic, flat GR case for \GCph+WL (blue) vs. \GCph+WL+XC (red) for the IA parameters and the 4 bias parameters, using the same galaxy bias model as above.}
\label{fig:triplot-flagship}
\end{center}
\end{figure*}

\autoref{tab: cosmoparflagflatgr} contains the $68\%$ forecast uncertainties on the cosmological parameters when the bias is described following the Flagship parameterisation. \autoref{fig:triplot-flagship} shows the improvement brought by XC, in the optimistic case, for the Flagship bias on cosmological parameters (left panel) and on the galaxy bias and IA parameters constraints (right panel).

\begin{table*}
\centering
\caption{$68\%$ errors on the 10 bias parameters of the binned bias model without and with the XC contribution. We consider a flat general relativity model, and set $\{\ell_{\rm max}({\rm WL}), \ell_{\rm max}($\GCph$)\}$ according to the pessimistic and optimistic assumptions.}
\resizebox{2\columnwidth}{!}{%
\begin{tabular}{cccccccccccc}
\hline
 \multicolumn{12}{c}{Bias parameters: flat cosmology, baseline}\\
\hline
Probe & Case & $\sigma(b_1)$ & $\sigma(b_2)$ & $\sigma(b_3)$ & $\sigma(b_4)$ & $\sigma(b_5)$ & $\sigma(b_6)$ & $\sigma(b_7)$ & $\sigma(b_8)$ & $\sigma(b_9)$ & $\sigma(b_{10})$ \\
\hline
 & pessimistic & 0.0075 & 0.0092 & 0.0097 & 0.0104 & 0.0112 & 0.0119 & 0.0127 & 0.0136 & 0.0152 & 0.0184 \\
WL+\GCph\ & & & & & & & & & & &\\
 & optimistic  & 0.0037 & 0.0045 & 0.0046 & 0.0049 & 0.0054 & 0.0057 & 0.0061 & 0.0065 & 0.0071 & 0.0082 \\
\hline 
 & pessimistic & 0.0074 & 0.0085 & 0.0091 & 0.0099 & 0.0105 & 0.0111 & 0.0116 & 0.0123 & 0.0131 & 0.0151 \\
WL+\GCph+XC & & & & & & & & \\
 & optimistic  & 0.0024 & 0.0032 & 0.0036 & 0.0041 & 0.0045 & 0.0049 & 0.0053 & 0.0055 & 0.0060 & 0.0068 \\
\hline

\end{tabular}
}
\label{tab: biasbaseflatgr}
\end{table*}

\begin{table*}
\centering
\caption{$68\%$ errors on the 10 bias parameters of the binned bias model without and with the XC contribution. We consider a general relativity model, where we relax the flatness  assumption. We set $\{\ell_{\rm max}({\rm WL}), \ell_{\rm max}($\GCph$)\}$ according to the pessimistic and optimistic assumptions.}
\resizebox{2\columnwidth}{!}{%
\begin{tabular}{cccccccccccc}
\hline
 \multicolumn{12}{c}{Bias parameters: non-flat cosmology, baseline}\\
\hline
Probe & Case & $\sigma(b_1)$ & $\sigma(b_2)$ & $\sigma(b_3)$ & $\sigma(b_4)$ & $\sigma(b_5)$ & $\sigma(b_6)$ & $\sigma(b_7)$ & $\sigma(b_8)$ & $\sigma(b_9)$ & $\sigma(b_{10})$ \\
\hline
 & pessimistic & 0.0127 & 0.0106 & 0.0098 & 0.0108 & 0.0125 & 0.0146 & 0.0173 & 0.0202 & 0.0246 & 0.0329 \\
WL+\GCph\ & & & & & & & & & & &\\
 & optimistic  & 0.0041 & 0.0047 & 0.0047 & 0.0049 & 0.0055 & 0.0059 & 0.0066 & 0.0071 & 0.0081 & 0.0099 \\
\hline 
 & pessimistic & 0.0118 & 0.0097 & 0.0092 & 0.0101 & 0.0115 & 0.0131 & 0.0151 & 0.0173 & 0.0208 & 0.0274 \\
WL+\GCph+XC & & & & & & & & \\
 & optimistic  & 0.0034 & 0.0036 & 0.0037 & 0.0041 & 0.0047 & 0.0053 & 0.0061 & 0.0065 & 0.0077 & 0.0095 \\
\hline
\end{tabular}
}
\label{tab: biasbasenonflatgr}
\end{table*}

\begin{table*}
\centering
\caption{$68\%$ errors on the cosmological parameters of the binned bias model without and with the XC contribution. We consider a flat modified gravity model, with departures from general relativity given by the $\gamma$ parameterisation. We set $\{\ell_{\rm max}({\rm WL}), \ell_{\rm max}($\GCph$)\}$ according to the pessimistic and optimistic assumptions.}
\begin{tabular}{cccccccccc}
\hline
 \multicolumn{10}{c}{Cosmological parameters: flat modified gravity cosmology, baseline}\\
\hline
Probe & Case & $\sigma(\Omegam)$ & $\sigma(\Omegab)$ & $\sigma(w_0)$ & $\sigma(w_a)$ & $\sigma(h)$ & $\sigma(n_{\rm s})$ & $\sigma(\sigma_8)$ & $\sigma(\gamma)$ \\
\hline
      & pessimistic & 0.0114 & 0.0034 & 0.157 & 0.621 & 0.0217 & 0.0127 & 0.0121 & 0.078 \\
WL+\GCph\ & & & & & & & & \\
      & optimistic & 0.0051 & 0.0024 & 0.068 & 0.246 & 0.0145 & 0.0043 & 0.0040 & 0.024 \\
\hline 
      & pessimistic & 0.0035 & 0.0027 & 0.064 & 0.333 & 0.0199 & 0.0108 & 0.0055 & 0.050 \\
WL+\GCph+XC & & & & & & & & \\
      & optimistic & 0.0020 & 0.0023 & 0.035 & 0.169 & 0.0137 & 0.0038 & 0.0020 & 0.021 \\
\hline
\end{tabular}
\label{tab: cosmogammaflat}
\end{table*}

\begin{table*}
\centering
\caption{$68\%$ errors on the 10 bias parameters of the binned bias model without and with the XC contribution. We consider a flat modified gravity model, with departures from general relativity given by the $\gamma$ parameterisation. We set $\{\ell_{\rm max}({\rm WL}), \ell_{\rm max}($\GCph$)\}$ according to the pessimistic and optimistic assumptions.}
\resizebox{2\columnwidth}{!}{%
\begin{tabular}{cccccccccccc}
\hline
 \multicolumn{12}{c}{Bias parameters: flat modified gravity cosmology, baseline}\\
\hline
Probe & Case & $\sigma(b_1)$ & $\sigma(b_2)$ & $\sigma(b_3)$ & $\sigma(b_4)$ & $\sigma(b_5)$ & $\sigma(b_6)$ & $\sigma(b_7)$ & $\sigma(b_8)$ & $\sigma(b_9)$ & $\sigma(b_{10})$ \\
\hline
      & pessimistic & 0.0087 & 0.0109 & 0.0159 & 0.0206 & 0.0249 & 0.0290 & 0.0334 & 0.0378 & 0.0436 & 0.0542 \\
WL+\GCph\ & & & & & & & & & & &\\
      & optimistic & 0.0038 & 0.0060 & 0.0076 & 0.0092 & 0.0107 & 0.0120 & 0.0136 & 0.0147 & 0.0168 & 0.0202 \\
\hline
      & pessimistic & 0.0079 & 0.0091 & 0.0120 & 0.0149 & 0.0177 & 0.0203 & 0.0230 & 0.0258 & 0.0295 & 0.0363 \\
WL+\GCph+XC & & & & & & & & & & &\\
      & optimistic & 0.0024 & 0.0045 & 0.0064 & 0.0081 & 0.0096 & 0.0109 & 0.0125 & 0.0136 & 0.0156 & 0.0188 \\
\hline 
\end{tabular}
}
\label{tab: biasgammaflat}
\end{table*}

\begin{table*}
\centering
\caption{$68\%$ errors on the cosmological parameters, without and with the XC contribution, when the bias is modeled following the Flagship simulation. We consider a flat general relativity model, and we set $\{\ell_{\rm max}({\rm WL}), \ell_{\rm max}($\GCph$)\}$ according to the pessimistic and optimistic assumptions.}
\begin{tabular}{ccccccccc}
\hline
 \multicolumn{9}{c}{Cosmological parameters: flat cosmology, Flagship bias}\\
\hline
Probe & Case & $\sigma(\Omegam)$ & $\sigma(\Omegab)$ & $\sigma(w_0)$ & $\sigma(w_a)$ & $\sigma(h)$ & $\sigma(n_{\rm s})$ & $\sigma(\sigma_8)$ \\
\hline
      & pessimistic & 0.0089 & 0.0030 & 0.1178 & 0.4060 & 0.0206 & 0.0095 & 0.0082 \\
WL+\GCph\ & & & & & & & & \\
      & optimistic & 0.0040 & 0.0023 & 0.0519 & 0.1754 & 0.0133 & 0.0038 & 0.0034 \\
\hline 
      & pessimistic & 0.0034 & 0.0026 & 0.0400 & 0.1604 & 0.0189 & 0.0090 & 0.0039 \\
WL+\GCph+XC & & & & & & & & \\
      & optimistic & 0.0018 & 0.0022 & 0.0253 & 0.0934 & 0.0130 & 0.0036 & 0.0017 \\
\hline
\end{tabular}
\label{tab: cosmoparflagflatgr}
\end{table*}

\bibliographystyle{aa}
\bibliography{references.bib}

\end{document}